\documentclass[preprint,12pt,authoryear,review]{elsarticle}
\usepackage{graphicx}
\usepackage{amsbsy,amsmath,amsthm,latexsym,amssymb}
\usepackage{color,ulem}
\usepackage{verbatim} 
\usepackage{lineno,hyperref}
\modulolinenumbers[5]
\newcommand{\bc}{\begin{center}}
\newcommand{\ec}{\end{center}}
\newcommand{\bfr}{\begin{flushright}}
\newcommand{\efr}{\end{flushright}}

\newcommand{\be}{\begin{enumerate}}
\newcommand{\ee}{\end{enumerate}}
\newcommand{\bi}{\begin{itemize}}
\newcommand{\ei}{\end{itemize}}
\newcommand{\bd}{\begin{description}}
\newcommand{\ed}{\end{description}}

\newcommand{\eeq}{\end{equation}}
\newcommand{\bea}{\begin{eqnarray}}
\newcommand{\eea}{\end{eqnarray}}

\newcommand{\bfi}{\begin{figure}}
\newcommand{\efi}{\end{figure}}
\newcommand{\bay}{\begin{array}{l}}
\newcommand{\eay}{\end{array}}

\newcommand{\cref}[1]{(\ref{#1})}   %to make cross reference easy.

\newcommand{\ie}{\textit{i.e.}~}
\newcommand{\ca}{\textit{ca.}~}
\newcommand{\cf}{\textit{cf.}~}
\newcommand{\eg}{\textit{e.g.}~}
\newcommand{\via}{\textit{via}~}
\newcommand{\viz}{\textit{viz.}~}
\newcommand{\vs}{\textit{vs.}~}
\newcommand{\eqname}{Eq.~}

\newcommand{\eqnames}{Eqs.~}
\newcommand{\figname}{Fig.~}

\newcommand{\fignames}{Figs.~}

\newcommand{\secname}{Section~}

\journal{Published article DOI:10.1061/JENMDT.EMENG-7896 under \copyright American Society of Civil Engineers.}

\begin{document}

\begin{frontmatter}

\title{MASKE: A kinetic simulator of coupled chemical and mechanical processes driving microstructural evolution}

\author{Enrico Masoero\corref{cor1}}

\cortext[cor1]{Corresponding author: enrico.masoero@polimi.it}

\affiliation{organization={Department of Civil and Environmental Engineering, Politecnico di Milano},%Department and Organization
            addressline={Piazza Leonardo da Vinci 32}, 
            city={Milan},
            postcode={20133}, 
            state={MI},
            country={Italy}}

\begin{abstract}
The microstructure of materials evolves through chemical reactions and mechanical stress, often strongly coupled in phenomena such as pressure solution or crystallization pressure. This article presents MASKE: a simulator to address the challenge of modelling coupled chemo-mechanical processes in microstructures. MASKE represents solid phases as agglomerations of particles whose off-lattice displacements generate mechanical stress through interaction potentials. Particle precipitation and dissolution are sampled using Kinetic Monte Carlo, with original reaction rate equations derived from Transition State Theory and featuring contributions from mechanical interactions. Molecules in solution around the solid are modelled implicitly, through concentrations that change during microstructural evolution and define the saturation indexes for user-defined chemical reactions. The structure and implementation of the software are explained first. Then, two examples on a nanocrystal of calcium hydroxide address its chemical equilibrium and its mechanical response under a range of imposed strain rates, involving stress-driven dissolution and recrystallization. These examples highlight MASKE's distinctive ability to simulate strongly coupled chemo-mechanical processes. MASKE is available, open-source, on GitHub.
\end{abstract}

%%Graphical abstract
%\begin{graphicalabstract}
%It's optional... skipping it for now
%\end{graphicalabstract}

%%Research highlights
\begin{highlights}
\item MASKE is presented: off-lattice Kinetic Monte Carlo on discretized microstructures
\item Original rate equations hard-coupling chemical reactions with mechanical stress
\item Dissolution/growth of a calcium hydroxide nanocrystal is simulated over long time
\item The crystal is put under realistic strain rates to compute stress-strain curves
\item A complex balance between solution chemistry, pressure solution, and plastic flow
\end{highlights}

\begin{keyword}
A. chemo-mechanical processes \sep C. numerical algorithms \sep A. microstructures \sep Kinetic Monte Carlo.
%% keywords here, in the form: keyword \sep keyword
%% PACS codes here, in the form: \PACS code \sep code
%% MSC codes here, in the form: \MSC code \sep code
%% or \MSC[2008] code \sep code (2000 is the default)
\end{keyword}

\end{frontmatter}

\section{Introduction}\label{secIntro}
% ---------------------------
% ---------------------------

A strong coupling between mechanical stress and chemical reactions often underlies the microstructural evolution of materials. Examples of stress-driven nucleation and growth of materials are commonly found in nature, \eg rock metamorphism or the adaptive growth of bones \citep{hart2017mechanical} and plants \citep{kouhen2023course}, but such processes are also exploited in nano-manufacturing, \eg the heterogeneous precipitation of islands on top of mismatching crystals, which is a challenge in semiconductor production \citep{dubrovskii2010stress}. Further examples concern the degradation of materials, such as crystallization pressure \citep{scherer2004stress}, pressure solution (\viz pressure-induced dissolution, \cite{rutter1983pressure}), creep from the relaxation of eigenstress originated during the formation of concrete \citep{bazant1997microprestress}, and stress-corrosion cracking \citep{raja2011stress}.

In Solid Mechanics, the modelling of chemo-mechanical processes is quite established at the macroscopic continuum level, with fully coupled formulations of reactive transport and linear momentum conservation, \eg in \cite{gawin2009modeling}. At the constitutive level, these formulations sometimes consider the stress or strain stemming from phase transformations, \eg local expansion during alkali-silica reaction and sulphate attack in concrete \citep{pesavento2012modeling,cefis2017chemo}, whereas the rates of chemical transformation are rarely stress-dependent \citep{bellomo2012constitutive}. Chemo-mechanical modelling is less advanced at the smaller microstructural level, where the focus in on the evolving morphology of single, or multiple, crystalline and amorphous phases. The length and time scales of such microstructural processes are too large for direct atomistic simulations, hence appropriate coarse-graining approaches are needed. However, most current simulations of microstructural evolution either describe the geometry of growing/dissolving solids without addressing their stress field, or they focus on the mechanical interactions prompting structural aggregation and collapse, but considering chemical reactions only qualitatively. Among the former, with focus on chemistry, there are microstructural development codes such as in  \cite{van1995numerical,bishnoi2009muic,bullard2011coupling}, as well as others simulators based on cellular automata \citep{bentz1994cellular,bullard2010parallel} or on Kinetic Monte Carlo (KMC) \citep{martin2020kimera}. Examples of simulators with a focus on mechanics are instead those in \cite{ioannidou2014controlling,ioannidou2016mesoscale,ioannidou2022review}, based on interacting particles. Exceptions exist, but with limitations. For example, \citep{li2015modeling} and \citep{feng2017multiscale} staggered finite element simulations of mechanical response with a reactive module to simulate solid precipitation, respectively of stress-free cement hydrates during creep and of expanding ettringite during sulphate attack in a cement paste. In both cases, however, chemo-mechanical coupling was not intrinsic to the model, but imposed by alternating chemical and mechanical modules. Differently, \cite{meakin2008simple} employed stress-dependent reaction rates in KMC simulations of crystal dissolution near stress-inducing dislocations. However the solid in these simulations was bound to a fixed lattice geometry (a cubic Kossel crystal), hence the stress field near the dislocation was pre-imposed and non-relaxing during dissolution. Another noteworthy method is the Phase Field, which does offer a framework to fully couple chemistry and mechanics; however, its high computational cost is preventing application to complex 3D structures, with multiple phases, liquid and solid, undergoing nonlinear mechanical processes \citep{tourret2022phase}.

This work presents MASKE: a simulator of fluid-solid microstructures discretized \via mechanically interacting particles whose dissolution and precipitation are sampled using off-lattice Kinetic Monte Carlo. MASKE allows the user to define multiple chemical reactions, along with their usual thermodynamic and kinetic parameters such as equilibrium constants and activation energies. Chemo-mechanical coupling is introduced through rate equations for the chemical reactions that feature the change in mechanical interaction energy accompanying particle dissolution or precipitation. Precursors of these rate equations where used in previous works. \cite{shvab2017precipitation} captured the experimental hydration rate curve of a cement paste by simulating the nucleation and stress-driven agglomeration of calcium-silicate-hydrate nanoparticles. \cite{coopamootoo2020simulations} simulated the experimentally measured critical undersaturation of a tricalcim silicate (C3S) paste, below which stress-driven stepwave dissolution is initiated at screw dislocations; recently, \cite{coopamootoo2023simulations} have extended the result to finite-sized crystal, capturing the full experimental relationship between dissolution rate and saturation index of C3S. \cite{alex2023carbonation} applied the coupled rate equations to the carbonation of a model cement paste, featuring multiple solid phases, ions in solution, and chemical reactions. These microscale simulations provided apparent rate constants that, employed in macroscale simulations of carbonation, matched a range of experimental results \citep{freeman2019indicator}. All these applications showed that the coupled reaction rates in MASKE can capture chemo-mechanical processes at experimentally realistic length and time scales. However, a detailed presentation and discussion of the architecture, capabilities, and limitations of MASKE are still due.

The algorithm and software architecture of MASKE are described in detail, accompanying its recent release, open-source, on GitHub \citep{maskecode}. The main article focuses on the most important parts, concept,s and steps in the algorithm; much more details are provided in several appendices, which guide the interested reader to a deeper understanding of the method, while also describing the input, output, and some data structures of the software. In the main article, particular attention is paid to: (i) the stress-dependent rates of chemical reaction implemented in MASKE, and their relationship with Transition State Theory (TST) and its predictions; (ii) the KMC algorithm to sample particle dissolution and off-lattice precipitation; (iii) the possibility to integrate continuous kinetic processes in time along with discrete KMC events. These key features are demonstrated \via two sample applications, both considering a spheroidal nano-crystal of calcium hydroxide Ca(OH)$_2$ in an aqueous solution of its ions. The chemical kinetics of Ca(OH)$_2$ under mechanical stress is of practical importance, as the instability and dissolution of this phase is responsible for pH drop and steel corrosion in reinforced concrete, and contribute to carbonation creep in cement pastes \citep{parrott1975increase}. Here however the focus is not on experimental validation, which has been already addressed in part by the applications summarised in the previous paragraph. Instead, the aim here is to showcase the three key features of MASKE listed above. Hence, the first application considers the chemical equilibrium, dissolution, and growth of the Ca(OH)$_2$ crystal in stress-free conditions; the results are compared with expectations from classical TST, and clarify the effect of some parameters in the rate equations and in the off-lattice sampling of precipitation. The second example puts the crystal under compression for a wide range of imposed strain rates; the results show that the interplay between mechanical deformations and stress-driven dissolution and recrystallization trigger rich and sometimes counter-intuitive mechanical responses, which are nevertheless explained by analysing the evolution of local stresses during the tests.

\section{Methodology}\label{secMethod}
% ---------------------------
% ---------------------------

This section presents for the first time the main structure and salient features of MASKE: a C++, object-oriented, parallel software that combines continuous processes, which are explicitly integrated in time, with discrete events, such as dissolution/precipitation reactions, sampled using off-lattice Kinetic Monte Carlo (KMC). In the Appendices, the interested reader can find full details of the MASKE algorithm, its implementation, the commands in its input script, and the parameters in its chemical database input file. In this section, attention is also paid to mechanical interactions between particles that are later used in two sample applications on Ca(OH)$_2$ nanocrystals. The methodological details of these applications are provided here too.

% ---------------------------
\subsection{Model description}
% ---------------------------

\begin{figure}[h]
\centerline{\includegraphics[width=0.45\textwidth] {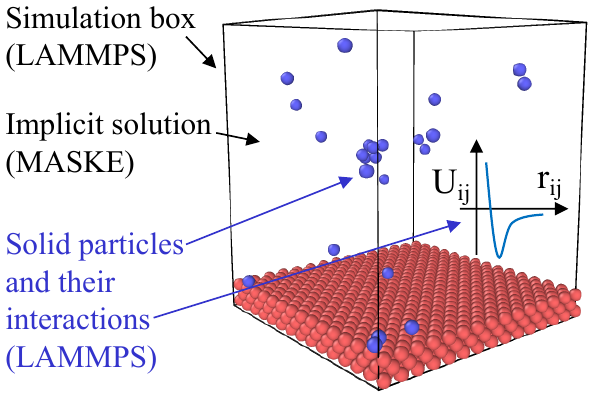}}
\caption{A typical simulation box in MASKE, with an implicit solution and solid particles interacing \via effective potentials (in this example, pairwise energy $U_{ij}$ as a function of interparticle distance $r_{ij}$).}
\label{figSystem}
\end{figure}

A typical simulation features spherical particles in a prismatic simulation box with volume $V_{box}$ and periodic or fixed boundary conditions: see \figname\ref{figSystem}. Box and particles are generated and tracked using the LAMMPS C++ library interface \citep{thompson2022lammps}. The particles may represent any explicit phase, hereafter assumed to be solid ones. The particles interact mechanically with each other through user-defined interaction potentials, which are also defined and managed in LAMMPS: \eg the $U_{ij}(r_{ij})$ interaction in \figname\ref{figSystem}. The current version of MASKE can only manage pairwise interactions, but extensions to any of the potentials in LAMMPS would be quite straightforward: see \ref{secAppA} for more discussion on this. Particle displacements caused by the mechanical interactions are also computed in LAMMPS, \via usual methods such as molecular dynamics or energy minimization (respectively using the $run$ and $minimize$ commands in LAMMPS). This interface with LAMMPS gives the user flexibility to study complex systems with various mechanical interaction potentials and loading conditions.

\begin{figure}[h]
\centerline{\includegraphics[width=0.45\textwidth] {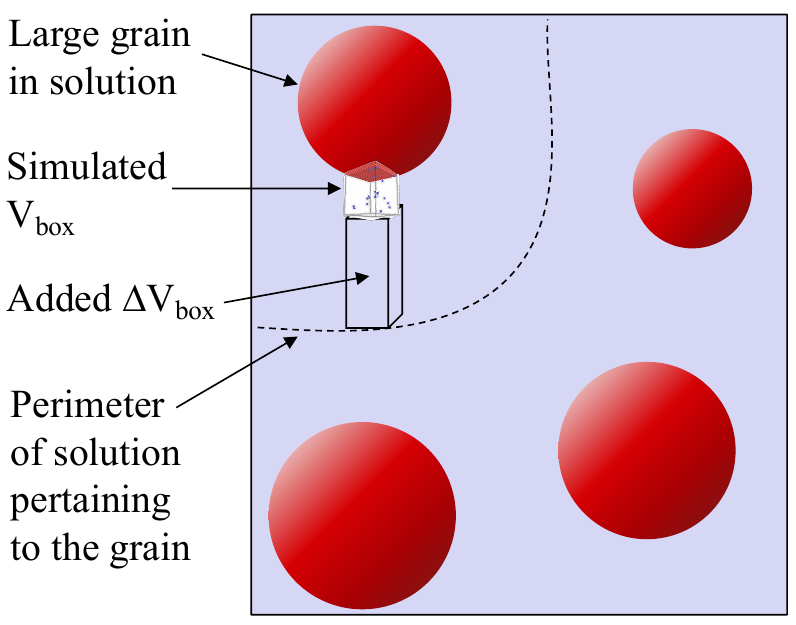}}
\caption{Rationale of adding a solution volume $\Delta V_{box}$, connected to the simulation box of volume $V_{box}$.}
\label{figDeltaV}
\end{figure}

The portion of a simulation box that is not occupied by solid particles contains a volume $voidV$ of empty space, and a volume $V_{sol}$ of implicit solution made up of various, user-defined molecular species. The current version of MASKE assumes that the molar concentrations of the solvated species are uniform in $V_{sol}$, with no diffusion taking place. There is also an additional volume $\Delta V_{box}$ attached to the simulation box, which can receive some of the molecules released during dissolution reactions, or provide some of the molecules consumed during precipitation reactions. This $\Delta V_{box}$ can be used to keep $V_{box}$ limited, which is where time-consuming operations take place; for example, \figname\ref{figDeltaV} depicts a scenario where this approach allows focussing on dissolution and precipitation near the surface of a large grain, while preserving realistic concentrations in solution and surface-to-volume ratio of suspended grains at a larger scale. The initial molar concentrations of all the solvated molecules in $V_{box}$ and in $\Delta V_{box}$ are provided by the user in the input script, along with fixing the temperature of the solution and the $A$ and $B$ solvent parameters to be used in Debye-H\"{u}ckel calculations of activity coefficients.

MASKE simulates discrete events, which happen once every $\Delta t_k$ units of time (where $\Delta t_k$ varies during a simulation depending on mechanical interactions and solution composition), but also continuous processes that are numerically integrated over a user-specified time interval $\Delta t_i$ with time step $dt_i$ (for the $i^{th}$ continuous process). In the current version of MASKE, the possible discrete events are full dissolution (called $deletion$) of any existing particle, or full precipitation (called $nucleation$) of a new particle at any possible location in $V_{box}$. The only continuous process implemented so far is called $CmstoreLMP$: it repeatedly  executes a set of stored LAMMPS commands, specified by the user in the MASKE input script (see \ref{secStructApp} for more details on the input/output structure of the code). A $CmstoreLMP$ process will be used in an example here, to impose a constant strain rate. The occurrence of discrete events and continuous processes is staggered onto a unique time line, $t^*$, which is advanced by whichever event or process has the shortest time of future occurrence, or $FOT$. The details of how this unique time line is managed can be found in \ref{secTimeApp}. The occurrence of discrete events is sampled using  Kinetic Monte Carlo (KMC), thus the list of all possible discrete events is associated with a unique $FOT_k$, which is the future occurrence time of any one out of all the possible discrete events. If $FOT_k$ is smaller than $FOT_i$ of all the other continuous processes, \ie if a discrete event is next in line to be executed, then one discrete event is picked out from the list of all the possible ones, with a probability that is proportional to its individual rate. This is the usual approach in KMC simulations; specifically, MASKE uses a rejection-free KMC algorithm that is time-dependent \citep{prados1997dynamical}, to account for changes in solution composition or stress state that continuous processes may induce between two successive discrete events. The equations of time-dependent KMC are also presented and discussed in \ref{secTimeApp}.

In the KMC method, $FOT_k$ is obtained from the cumulative rate:
\begin{equation}
R_k = \sum_j^{N+M} r_{k,j} 
\end{equation}
which is the sum of the individual rates $r_{k,i}$ of all $N$ possible particle deletions plus all $M$ particle nucleations at all possible locations in the simulation box. The expressions for $r_{k,j}$ are presented in the next section. $N$ is finite, whereas the number of possible locations for particle precipitation is infinite and, in principle, one should sample them all to use KMC. To treat this problem, MASKE asks the user to select a region within $V_{box}$ and to create a 3D lattice of $M$ trial particles of the desired type, \viz with desired chemical composition and interaction potentials. Each of these $M$ trial particles is then driven towards its closest local minimum of interaction energy with the other already-existing, so-called $real$ particles (no interactions between trial particles are considered, and the positions of the real particles are kept fixed while moving the trial ones). This energy minimization is performed using a bespoke algorithm, called $quickmaske$, which the $quickmin$ algorithm in LAMMPS except that each trial particles is independently relaxed. The fineness of the user-specified lattice determines the volume $\Delta V$ of the generic lattice cell. The energy minimum obtained for a trial particle is assumed to characterize all the particles that may nucleate anywhere else in $\Delta V$. In this way, the cumulative rate of all the $M$ sampled nucleation events is effectively a numerical integral in the sampled region, with resolution $\Delta V$. The most accurate result is obtained when $\Delta V \rightarrow 0$, but in practice a small but finite $\Delta V$ is already sufficient to estimate the cumulative nucleation rate accurately. How small $\Delta V$ should be, it depends on the employed interaction potentials and on the morphology of the simulated microstructure. An assumption underling this approach is that the energy minima alone fully control the precipitation rate cumulated over the whole $V_{box}$. In reality, thermal fluctuations would add contributions also from the locations surrounding the minima; future versions of MASKE may sample such fluctuations by relaxing the trial particles using molecular dynamics instead of energy minimization. \ref{secSampleApp} offers more discussion on interaction energy landscapes and their sampling for dissolution and precipitation events.

\begin{figure}[h]
\centerline{\includegraphics[width=0.99\textwidth] {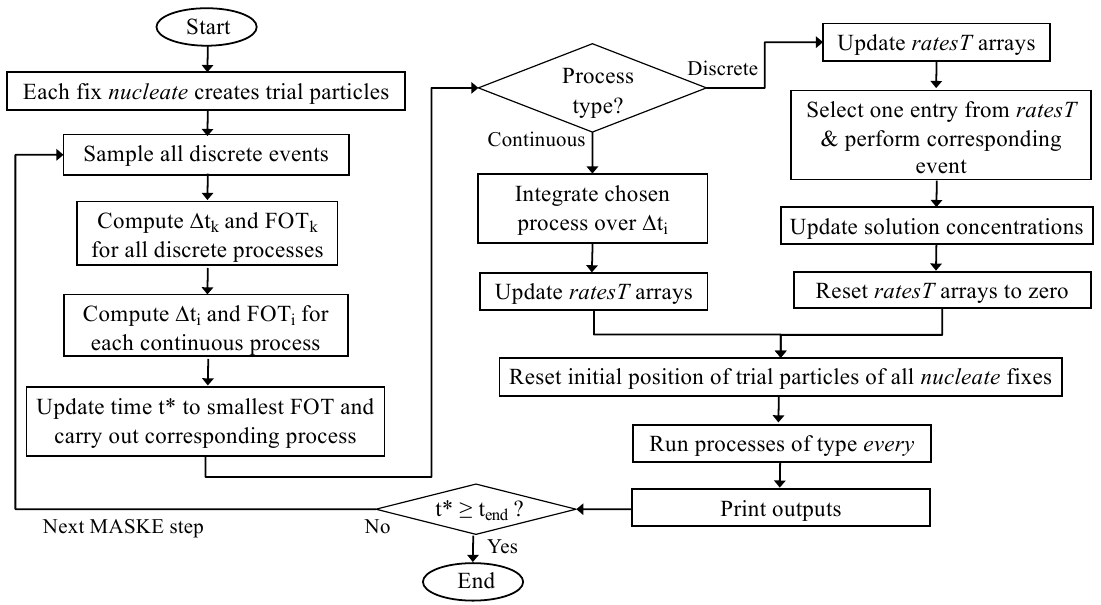}}
\caption{Flow chart of the main kinetic loop in MASKE.}
\label{figKrunFlow}
\end{figure}

The flowchart in \figname\ref{figKrunFlow} depicts the algorithm for the kinetic simulations in MASKE. A brief description is given here, while more details can be found in \ref{secAppkrun}. The left side of the flowchart summarises the sampling of $FOT$ for all the discrete and continuous processes in a simulation. The shortest $FOT$ determines which type of process or event is to be executed next. If a continuous process is next, it is directly integrated numerically over its time period $\Delta t_i$ (user-specified in the input script). After that, the individual rates $r_{k,j}$ of the sampled discrete events are multiplied times $\Delta t_i$ and these products are stored in a per-particle array called $ratesT$, of size $N+M$. The $ratesT$ array is used in the next kinetic step for computing $FOT_k$. If the next event to execute is instead a discrete one, the $ratesT$ values are immediately updated, adding to each entry the product $r_{k,j} (FOT_k-t^*)$. The $ratesT$ values are then used to select which discrete event to carry out specifically, out of all the possible ones. After executing a discrete event, the concentrations of the solvated molecules are updated accordingly, as discussed in the next section, and the $ratesT$ values are reset to zero. Subsequently, irrespective of whether the executed process is continuous or discrete, MASKE restores the initial position of all trial particles and executes a third type of possible processes, called $Every$. Unlike continuous or discrete processes, whose occurrence is associated to a time line, processes of type $Every$ are simply executed once every $n$ kinetic steps, where a step is one  full loop in \figname\ref{figKrunFlow}. $n$ is user-specified as an input, and presently MASKE implements only one type of $Every$ process, called $EmstoreLMP$, which executes a set of LAMMPS command that the user can specify and store in the input script. A typical use of such an $Every$ process is to execute a full energy minimization of the system with $n=1$, \ie after each execution of a continuous or discrete process.

The algorithm in \figname\ref{figKrunFlow} is well suited to parallelisation, since each discrete event (\ie each of the $N$ particle deletions and $M$ particle nucleations) can be sampled independently. Furthermore, spatial domain decomposition can be exploited during energy minimization. To exploit this, MASKE features a two-level parallelisation. On the first level, the user can specify multiple groups of processors, called $subcommunicators$, and attribute the sampling of one or more processes to each subcommunicator (\eg one subcommunicator samples particle deletion, another one samples nucleation). On the second level, each subcommunicator uses all its processors to run any invoked LAMMPS command, \eg energy calculations and minimisation using domain decomposition. This article does not report on the scalability and efficiency of the parallelisation, which is amenable to optimisation in the current version of MASKE. However, previously published works have already exploited both levels of parallelisation to simulate hundreds of thousands particles \citep{coopamootoo2020simulations} and multiple types of solid particles, all sampled for deletion and nucleation \citep{alex2023carbonation}.

\subsection{Chemical reactions: rates and mechanisms}

MASKE computes the rate of any discrete event, such as particle deletion or nucleation, based on two user-defined entities: a chemical reaction and a reaction mechanism, both specified in a chemical database text file, called $ChemDB$, whose syntax is presented in \ref{secChemDBApp}. The chemical reaction is assumed to be one-step and is defined primarily through its formula and stoichiometric coefficients, \eg:
\begin{align}
Ca(OH)_2 \rightarrow Ca^{2+} + 2 OH^-   \label{eqStoi}
\end{align}
for the dissolution of a calcium hydroxide molecule. The stoichiometric coefficients are specified in the $ChemDB$ file. The molecules appearing in the reaction formulas are also user-defined in the $ChemDB$ file, along with a set of useful parameters. For the $i^{th}$ solvated molecule, these parameters are its apparent volume $v_{app,i}$, hydrated radius $a^o_i$, ionic charge $z_i$, and the $b_i$ parameter for Debye-H\"{u}ckel calculations of activity coefficients, discussed later. For solid molecular species, only linear sizes are required, which are used to compute their molecular volumes. $ChemDB$ gathers also the thermodynamic and kinetic parameters associated to each chemical reaction: standard free energy barriers $\Delta G^{o\ddag}$, standard concentrations of activated complexes $c^{o\ddag}$ (in the sense of Transition State Theory, which will be used below to express rate equations), equilibrium constants $K_{eq}$, and interfacial energies $\gamma$. See \ref{secChemDBApp} for more details on this quantities.

The $ChemDB$ file also contains the definition of reaction mechanisms, which are the sets of assumptions that MASKE makes when assembling single chemical reactions to delete or nucleate a full particle. Indeed, a particle may be a coarse-grained entity whose deletion or creation may entail more than just a single reaction, \eg $n_{r,j,diss}$ or $n_{r,j,nuc}$ reactions respectively to dissolve or nucleate a particle. Classical Nucleation Theory is an example of multiple individual reactions being assembled to express a reaction rate at the particle scale. In the current version of MASKE, the only implemented mechanism is called $allpar$ and it assumes that, irrespective of how big a particle is, its full deletion or nucleation only takes the time for one single-step chemical reaction to occur. This is exact if the volume $V_{p,j}$ of the generic particle $j$ equals the sum of the volumes of solid molecules dissolving or precipitating in one reaction. For bigger particles instead, the $allpar$ mechanism effectively assumes that the same chemical reaction takes places everywhere in $V_{p,j}$ at the same time. This is clearly unrealistic for particles beyond the nanometre or so, therefore in this article all the examples will feature particles whose $V_{p,j}$ entail $n_{r,j,diss} = n_{r,j,nuc} = 1$. Hereafter this resolution is called 'unimolecular', although in principle a single chemical reaction might also involve multiple solid molecules at once.

Whenever a particle is deleted or nucleated, a chemical reaction is carried out as part of an $allpar$ mechanisms. MASKE ensures mass conservation by consistently managing molecule exchanges between solid phases and solution, following the user-specified reaction formulas in $ChemDB$. Chemical reactions may entail differences in volume between reactants and products; these are accommodated in the $voidV$ volume introduced in the previous section, which may thus be positive or negative. The current version of MASKE does not simulate chemical reactions taking place in the implicit solution, hence it does not minimize the free energy of the solution itself. An interface between MASKE and the thermodynamic simulator PHREEQC \citep{parkhurst1995user} is currently being developed, to address this limitation.

The rate equations for the $allpar$ mechanis in MASKE are rooted in Transition State Theory (TST), but with an extension to account for excess enthalpy coming from the mechanical interactions between particles. This excess enthalpy term, originally proposed in \cite{shvab2017precipitation}, is essential for capturing the effect of local morphology and stress state on reaction rates. The rate equations for one-step dissolution and precipitation reactions are:
\begin{align}
r_{1,d} &= \kappa \frac{k_BT}{h}\frac{c^{o\ddag}_d}{\gamma^\ddag_d} \exp\left( \frac{-\Delta G^{o\ddag}_{d}}{k_BT} \right) \exp\left[ -(1-\chi)\frac{\Delta U_{d}-\gamma \Omega}{k_BT} \cdot \frac{1}{n_{r,j,diss}} \right] Q_{r,d} V_t^{\alpha/3}  \label{eqr1d} \\
r_{1,pr} &= \kappa \frac{k_BT}{h}\frac{c^{o\ddag}_{pr}}{\gamma^\ddag_{pr}} \exp\left( \frac{-\Delta G^{o\ddag}_{pr}}{k_BT} \right) \exp\left[ -\chi\frac{\Delta U_{pr}+\gamma \Omega}{k_BT} \cdot \frac{1}{n_{r,j,nuc}}\right] Q_{r,pr} \Delta V \cdot V_t^{\alpha/3-1}  \label{eqr1pr}
\end{align}
\eqnames\ref{eqr1d} and \ref{eqr1pr} will be hereafter called 'straight' rate equations, as opposed to 'net' rates defined later in this section. $\kappa$ is the probability of an activated complex evolving into reaction products; MASKE makes the common assumption that $\kappa = 1$ \citep{lasaga2014kinetic}. $k_B$ and $h$ are the Boltzmann and Planck constants, and $T$ is the temperature in degrees Kelvin. $c^{o\ddag}_d$ and $c^{o\ddag}_{pr}$ are the standard concentrations of activated complexes for the dissolution and precipitation reactions, $\gamma^{\ddag}_d$ and $\gamma^{o\ddag}_{pr}$ are their activity coefficients, and $\Delta G^{o\ddag}_d$ and $\Delta G^{o\ddag}_{pr}$ are their standard free energies of activation. $c^{o\ddag}$ and $\Delta G^{o\ddag}$ can be estimated experimentally, noting that the whole prefactors in \eqnames\ref{eqr1d} and \ref{eqr1pr} defines the rate constant $k$:
\begin{equation}
k = \kappa \frac{k_BT}{h}\frac{c^{o\ddag}}{\gamma^\ddag_d} \exp\left( \frac{-\Delta G^{o\ddag}}{k_BT} \right) \label{eqkconst}
\end{equation}
\eqname\ref{eqkconst} implies that the arbitrary choice of $c^{o\ddag}$ must be compensated by the value of $\Delta G^{o\ddag}$, so that a physical, univocally defined rate constant is preserved. The details of this compensation are discussed in \ref{secSampleApp}.

The $V_t$ term in \eqnames\ref{eqr1d} and \ref{eqr1pr} is the tributary volume of the particle being sampled for deletion or nucleation, \viz the volume of the interaction energy basin, whose local energy minimum is where the sampled particle sits. The user of MASKE must specify $V_t$ in the $ChemDB$ file; for solid phases governed by short-range interactions, $V_t$ is typically in the range of one particle volume (see \ref{secSampleApp} for more discussion and examples of $V_t$). $\alpha$ is the spatial dimensionality of $c^{o\ddag}_d$ or $c^{o\ddag}_{pr}$; namely, $\alpha = 3$ if the reactions are per unit volume or, as typical for dissolution/precipitation, $\alpha = 2$ indicating reactions per unit surface. $\Delta V$ is the cell volume of the user-defined lattice of trial particles sampling candidate nucleation sites. When the KMC algorithm sums all the nucleartion rates for trial particles pertaining to the same interaction energy basin, the sum of all $\Delta V$'s equals $V_t$ (within the limits posed by the finite, and not infinitesimal, value of $\Delta V$). As a result, $V_t^{\alpha/3}$ in \eqnames\ref{eqr1d} equals the sum of all $\Delta V \cdot V_t^{\alpha/3-1}$ pertaining to the same interaction energy basin. This aspect is discussed in more detail in \ref{secAppNetTST}.

The square brackets in \eqnames\ref{eqr1d} and \ref{eqr1pr} quantify the excess enthalpy of a particle, resulting from two contributions: $\Delta U_d$, which is the change in total interaction energy in the system caused by the particle deletion ($\Delta U_{pr}$ is the analogous term for a nucleation even), and$\gamma$, which is the solid-solution interfacial energy attributed to the sampled particle, and user-defined in the $ChemDB$ file. The $\Omega$ term, which multiplies $\gamma$ in the excess enthalpy, is the surface area of the sampled particle. $n_{r,j,diss}$ and $n_{r,j,nuc}$ appear in the excess enthalpy because the $allpar$ mechanism may be used for particles whose deletion or nucleation entails multiple chemical reactions; for the unimolecular resolution in this article, $n_{r,j,diss} = n_{r,j,nuc} = 1$. The user-defined parameter $\chi$, with value between 0 and 0.5, specifies which fraction of excess enthalpy $U + \gamma \Omega$ is still present in the system when the reaction is in its activated complex state. $\chi = 0$ is the usual assumption in TST, but results in this article will show that $\chi>0$ may sometimes improve the efficiency of a simulation.

The last terms to be defined in the rate equations are $Q_{r,d}$ and $Q_{r,pr}$, which are the activity products of the reactants in the dissolution and precipitation reactions. If $i$ is the generic solvated reactant:
\begin{equation}
Q_r = \prod_{i} (\gamma_i c_i)^{\nu_i} \label{eqQr}
\end{equation}
where $c_i$ is the concentration in solution of the molecular species $i$, $\nu_i$ is its stoichiometric coefficient in the reaction, and $\gamma_i$ is its activity coefficient. MASKE computes $\gamma_i$ using the WATEQ Debye-H\"uckel formula \citep{truesdell1974wateq}:
\begin{equation}
log_{10} (\gamma_i) =  \frac{-z_i^2 A \sqrt{I}}{1+B a_i^o \sqrt{I} } + b_i I  \label{eqDebH}
\end{equation}
$z_i$ is the charge of the $i^{th}$ species. $A$ and $B$ are solvent-specific constants (0.51 and 3.29 nm$^{-1}$ for water). $a_i^o$ is the hydrated radius of the molecule in solution, and $b_i$ is a molecule-specific dimensionless constant; both these parameters are provided by the user in the $ChemDB$ file. $I$ is the ionic strength of the fluid, featuring $n_s$ solvated species:
\begin{equation}
I =  \sum_{j=1}^{n_s} c_j z_j^2 	\label{eqIstr}
\end{equation}
When $a_i^o$ or $b_i$ are unknown, MASKE computes the activity coefficients using established simplified methods \citep{langmuir1997aqueous}, detailed in \ref{secSampleApp}.

The straight rate expressions in \eqnames\ref{eqr1d} and \ref{eqr1pr} sample all the fluctuations between particle deletion and nucleation that, on average, guide the overall morphology evolution of the system. However, many such fluctuations may cancel out, especially near equilibrium, leading to inefficient simulations where the overall morphology makes very little progress on average. In such scenarios it may be convenient to remove the fluctuations by considering net rates instead, defined as the difference between straight rates: 
\begin{align}
r_{1,d}^{net} &= \max\langle 0,r_{1,d} - r_{1,pr}\rangle = \nonumber \\
& = \max\left\langle 0,  \kappa \frac{k_BT}{h}  \frac{c^{o\ddag}_d}{\gamma^\ddag_d} \exp\left( \frac{-\Delta G^{o\ddag}_{d}}{k_BT} \right) V_t^{\alpha/3} \cdot  \right. \nonumber \\
& \hspace{1.5cm} \left. \left\{  \exp\left[ -(1-\chi)\frac{\Delta U_{d}-\gamma \Omega}{k_BT} \cdot \frac{1}{n_{r,j,diss}} \right] Q_{r,d} -	 \exp\left[ -\chi\frac{-\Delta U_{d}+\gamma \Omega}{k_BT} \cdot \frac{1}{n_{r,j,diss}}\right] \frac{Q_{p,d}}{K_{eq,d}} \right\}  \right\rangle \label{eqrNd} \\
r_{1,pr}^{net} &= \max\langle 0,r_{1,pr} - r_{1,d}\rangle = \nonumber \\
& = \max\left\langle 0,  \kappa \frac{k_BT}{h}  \frac{c^{o\ddag}_{pr}}{\gamma^\ddag_{pr}} \exp\left( \frac{-\Delta G^{o\ddag}_{pr}}{k_BT} \right) \Delta V \cdot V_t^{\alpha/3-1} \cdot  \right. \nonumber \\
& \hspace{1.cm} \left. \left\{  \exp\left[ -\chi\frac{\Delta U_{pr}+\gamma \Omega}{k_BT} \cdot \frac{1}{n_{r,j,nuc}}\right] Q_{r,pr} - \exp\left[ -(1-\chi)\frac{-\Delta U_{pr}-\gamma \Omega}{k_BT} \cdot \frac{1}{n_{r,j,nuc}} \right] \frac{Q_{p,pr}}{ K_{eq,pr}} \right\}  \right\rangle \label{eqrNpr}
\end{align}
These net rate equations are also implemented in MASKE, as a possible option for the $allpar$ mechanism. Their derivation is detailed in \ref{secSampleApp}. $Q_{p,d}$ and $Q_{p,pr}$ are the activity products of the products of the dissolution and precipitation reactions, with $K_{eq,d}$ and $K_{eq,pr}$ their equilibrium constants. The reactants of a dissolution reaction and the products of a precipitation reaction are solid phases, conventionally in standard state when stress-free, thus $Q_{r,d} = Q_{p,pr} = 1$. \ref{secAppNetTST} shows how the net rates in \eqnames\ref{eqrNd} and \ref{eqrNpr} effectively scale linearly with the saturation index $\beta$ of the solution for the associated chemical reaction; namely, $r_{1,d}^{net}\sim(1-\beta)$ and $r_{1,pr}^{net}\sim(\beta-1)$, as usual in Transition State Theory. The definition of $\beta$ is:
\begin{equation}
\beta = \frac{Q_{r,pr}}{K_{eq,d}} = \frac{Q_{p,d}}{K_{eq,d}}
\end{equation}
noting that both the reactants of precipitation and the products of dissolution are simply the solvated species participating in the reaction. In a scenario where there is no excess enthalpy from changes in mechanical stress and interfacial energies, $\beta = 1$ would imply equilibrium, \ie no dissolution nor precipitation taking place, whereas $\beta > 1$ would favour precipitation and $\beta < 1$ dissolution.

% ---------------------------
\subsection{Interaction potentials between identical particles}
% ---------------------------

The energy $U$ from mechanical interactions plays an important role in the reaction rates in MASKE, \via the excess enthalpy terms $\Delta U_d-\gamma \Omega$ and $\Delta U_{pr}+\gamma \Omega$ in \eqnames\ref{eqr1d} and \ref{eqr1pr}, as implemented in the $allpar$ mechanism. All the examples in this article will employ a truncated harmonic potential of interaction between pairs of identical, spherical particles, $i$ and $j$:
\begin{align}
U_{ij} &= \frac{1}{2} k_0 \left( r_{ij} - r_0 \right)^2 - \varepsilon_0 \;\;\; \mathrm{if} \; r_{ij}<r_c  \nonumber  \\
U_{ij} &=   0  \;\;\; \mathrm{if} \; r_{ij} \ge r_c
\label{eqUij}
\end{align}
\begin{figure}[h]
\centerline{\includegraphics[width=0.45\textwidth] {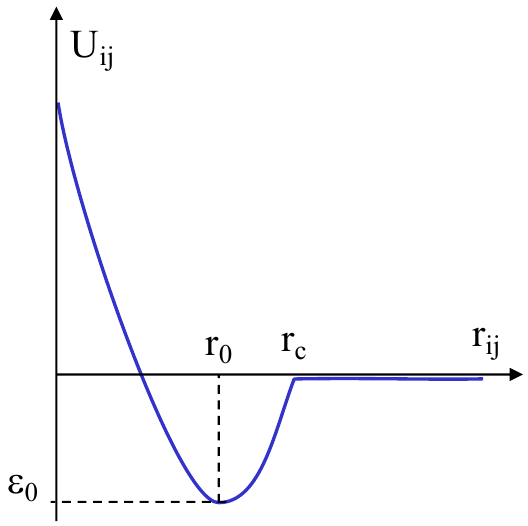}}
\caption{Truncated harmonic potential between two identical particles $i$ and $j$, with interparticle distance $r_{ij}$, equilibrium distance $r_0$ with minimum $U_{ij} = -\varepsilon_0$, and interaction cutoff $r_c$.}
\label{figUij}
\end{figure}
plotted in \figname\ref{figUij}. $r_{ij}$ is the interparticle distance. When the particles are at their equilibrium distance $r_0$, the interaction energy $U_{ij}$ is minimum and equal to the bond energy $-\varepsilon_0$. The stiffness of the interaction is dictated by the $k_0$ parameter. The cutoff distance for the interaction, $r_c$, is set here as the distance at which $U_{ij}$ returns to zero when $r_{ij} > r_0$; because of this definition, $r_c$ is not an independent parameter, but it is fixed once $r_0$, $k_0$, and $\varepsilon_0$ are decided:
\begin{equation}
r_c = r_0 + \sqrt{\frac{2\varepsilon_0}{k_0}}  \label{eqrc}
\end{equation}
$k_0$ is obtained from the Young modulus $E$ of the material of the particles:
\begin{equation}
k_0 = \pi\frac{E (D_0/2)^2}{r_0} \label{eqk0}
\end{equation}
The particle diameter $D_0$ is close but not equal to $r_0$, and their relationship depends on the modelling choice of the `limit crystal lattice' of the solid phase. Depending on the type and range of the employed interaction potential, there are only several ordered arrangements, \viz crystal lattices, that are mechanically stable and where all the interacting particles sit at equilibrium distances. For instance, limit crystal lattices for the short-ranged, first-neighbour-only, radial interactions in \figname\ref{figUij}, would be face centred cubic (FCC) or hexagonal close packing (HCP), both with $n_b = 12$ first neighbours in the bulk. The mismatch between $r_0$ and $D_0$ emerges because any limit crystal lattice of spherical particles features a porosity and fills the space only up to a volume fraction $\eta < 1$. For example, both FCC and HPC lattices have packing fraction $\eta = 0.74$, hence a 26\% porosity. However, if these limit lattices are used to discretise a non-porous continuum in a simulation, the maximum packing $\eta$ should represent a solid with same density as a non-porous phase, \viz with the same number of molecules, or mass, per unit volume. To exactly compensate for the artificial porosity $1-\eta$, \cite{coopamootoo2020simulations} have proposed to use an $r_0$ smaller than $D_0$ by:
\begin{equation}
r_0 = D \sqrt[3]{\eta} \label{eqr0}
\end{equation}

The expressions of excess enthalpy in \eqnames\ref{eqr1d} and \ref{eqr1pr} ($\Delta U \pm \gamma \Omega$) impose a dependence between interaction potential $\varepsilon_0$ and solid-solution interfacial energy $\gamma$, which must be respected in order to produce physically meaningful simulations. Consider a particle in kink position in the crystal lattice, which by definition has $n_b/2$ first neighbours. Precipitating or dissolving a stress-free particle in a kink position should not change the excess enthalpy of the system, because it creates the same amount of solid-solution interfacial area as it removes. Such a dissolution (and similarly, in reverse, for precipitation) can be modelled, first, as a detachment of the kink particle from the solid, which causes a loss of $n_b/2$ inter-particle bonds each with energy $\varepsilon_0$ (accounted for in the $\Delta U$ term), and second, as a dissolution of the detached particle which removes $\gamma \Omega$ interfacial energy. This leads to:
\begin{equation}
\varepsilon_0 \frac{n_b}{2} = \gamma \Omega \;\; \rightarrow \;\; \varepsilon_0 =  \frac{2 \gamma \Omega}{n_b}  \label{eqeps0}
\end{equation}
%

% ---------------------------
\subsection{Interaction potentials between different particles}
% ---------------------------

One example in this article will feature mechanical interactions between particles of two different types, with different sizes and representing different materials. For such scenarios, \cite{alex2023carbonation} have proposed an extension of the truncated harmonic potential in \figname\ref{figUij}:
\begin{align}
U_{ij} &= \frac{1}{2} k_{12} \left( r_{ij} - r_{12} \right)^2 - \varepsilon_{12} \;\;\; \mathrm{if} \; r_{ij}<r_{c,12}  \nonumber  \\
U_{ij} &=   0  \;\;\; \mathrm{if} \; r_{ij} \ge r_{c,12}
\label{eqUijDiss}
\end{align}
The equilibrium distance $r_{12} = (r_{0,1}+r_{0,2})/2$ is the average of the equilibrium distances between identical particles for the two different types, 1 and 2. The stiffness term is expressed as $k_{12} = 2\frac{k_1k_2}{k_1+k_2}$, \ie idealizing the interacting particles as two springs in series, with individual stiffness as per \eqname\ref{eqk0}. The bond energy is $\varepsilon_{12} = \frac{\gamma_{12}(\Omega_1+\Omega_2)}{n_b}$, with $\gamma_{12}$ the solid-solid interfacial energy between the particles and $\Omega_1$ and $\Omega_2$ their individual surface areas. \eqname\ref{eqrc} is still valid for the cutoff distance, using $r_{12}$, $\varepsilon_{12}$, and $k_{12}$ in it.

% ---------------------------
\subsection{Example 1: dissolution, growth, and equilibrium of a stress-free crystal}\label{secEx1}
% ---------------------------

A nanocrystal of calcium hydroxide, Ca(OH)$_2$, is considered in this example. The crystal is immersed in an aqueous solution of Ca$^{2+}$ and OH$^-$ ions at room temperature $T = 298$ K. The crystal is made of 4,055 spherical particles, each with the volume of one Ca(OH)$_2$ molecule, $V_{p,CH} = \frac{V_{M,CH}}{N_{a}} = 0.0557$ nm$^3$, where $N_a$ is the Avogadro number and $V_{M,CH} = 33.53$ cm$^3$/mol is the molar volume of calcium hydroxide (indicated as $CH$ in the subscript). This brings to a particle diameter $D_{CH} = 0.4737$ nm. The particles are initially arranged on an FCC lattice with cell size $a = \sqrt{2} \; r_{0,CH}$, which implies that $r_{0,CH}$ is the distance between first-neighbour particles in the lattice. Given the packing density $\eta = 0.74$ of the FCC lattice, \eqname\ref{eqr0} provides $r_{0,CH} = \sqrt[3]{\eta}\;D_{CH} = 0.4285$ nm.
\begin{figure}[h]
\centerline{\includegraphics[width=0.95\textwidth] {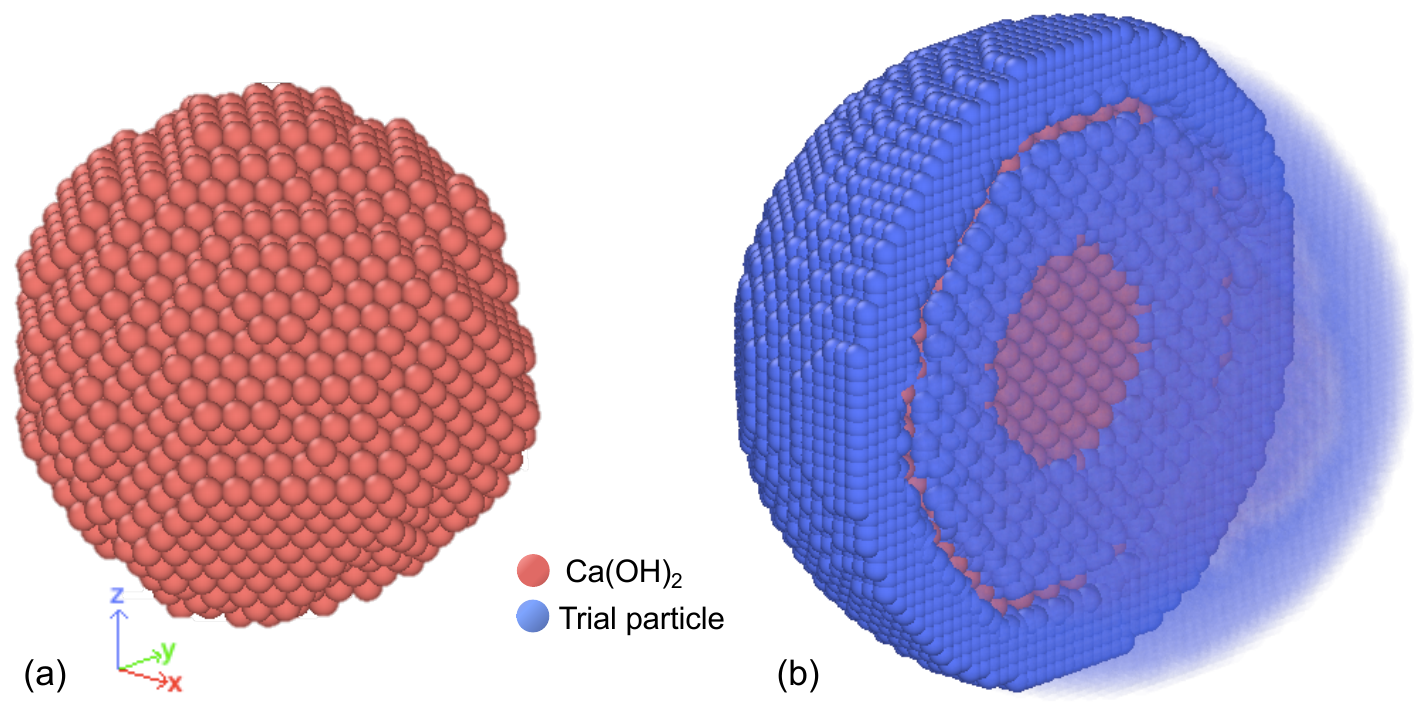}}
\caption{(a) Initial configuration of the Ca(OH)$_2$ nano-crystal, obtain from pre-dissolving a cubic crystal of Ca(OH)$_2$ particles in FCC arrangement; (b) A section of the same nano-crystal surrounded by trial particles, used to sample precipitation of new particles. The trial particles are shown after minimization to their closest local energy minima, hence those in the inner part of the spherical shell region leave the perfect cubic arrangement of the initial lattice, driven by mechanical interactions with the other, already existing Ca(OH)$_2$ particles (in red). Instead, the trial particles in the outer part of the spherical shell do not interact with any existing particle, hence they maintain the ordered cubic arrangement of the trial lattice.}
\label{figEx1init}
\end{figure}
The interaction potential between particles is the truncated harmonic one from \eqname\ref{eqUij}. From the Young modulus $E_{CH} =  48$ GPa \citep{wittmann1986estimation} of calcium hydroxide, \eqname\ref{eqk0} provides $k_{0,CH} = 19,745$ MPa$\cdot$nm. The bond energy $\varepsilon_{0,CH}$ comes from \eqname\ref{eqeps0}, having considered an interfacial energy between calcium hydroxide and surrounding solution $\gamma = 68.4$ mJ/m$^2$ \citep{estrela2005surface} and $n_b = 12$ first neighbours for a bulk particle in the FCC lattice. The interaction cutoff distance thus becomes $r_{c,CH} = 0.457$ nm, from \eqname\ref{eqrc}; this entails a failure strain $\frac{r_{c,CH}-r_{0,CH}}{r_{0,CH}} \approx 6.5\%$, which is reasonable at the molecular scale. With these interactions, an initially cubic grain of FCC-arranged calcium hydroxide particles is pre-dissolved as in \cite{masoero2023maske}, obtaining the rounded grain with diameter of \ca 7.3 nm in \figname\ref{figEx1init}, which is the configuration for this example. 

A very large volume of solution is placed in contact with the solid grain, by setting $\Delta V_{box} = 10^{20}$ nm$^3$. In this way, particle dissolution and precipitation negligibly change the concentration of solvated ions, which retain their initially assigned values \footnote{The apparent volumes of solvated Ca$^{2+}$ and OH$^-$ ions, set to -0.025129603 nm$^3$ and 0.001030404 nm$^3$ respectively \citep{lothenbach2019cemdata18} in the $ChemDB$ database, have a negligible effect in this example due to the large $\Delta V$ attached to the simulation box}. The chemical reactions underlying dissolution and precipitation here are: 
\begin{align}
Ca(OH)_2 \rightarrow Ca^{2+} + 2 OH^- \;\;\; &\mathrm{with} \; K_{eq,d} = 6.30866 \cdot 10^{-6}  \nonumber  \\
Ca^{2+} + 2 OH^- \rightarrow Ca(OH)_2 \;\;\; &\mathrm{with} \; K_{eq,pr} = \frac{1}{K_{eq,d}} = 1.58512 \cdot 10^5
\label{eqCHstoi}
\end{align}
The value of $K_{eq,d}$ is taken from \cite{lothenbach2019cemdata18}. The solvated ions are assigned the following parameters to calculate their activities as well as the ionic strength of the solution: $z_{Ca}=+2$, $a_{Ca}^o = 0.486$ nm, $b_{Ca}=0.15$, and $z_{OH}=-1$, $a_{OH}^o = 1.065$ nm, $b_{OH}=0.064$ \citep{lothenbach2019cemdata18}. In this way, the activity product of the solvated ions for \eqname\ref{eqCHstoi} can be computed for any initially imposed set of concentrations. The explored concentrations range from 0 for both Ca$^{2+}$ and $OH^-$ (saturation index $\beta = 0$) to $c_{Ca} = 39.1$ mmol/L, $c_{OH} = 78.2$ mmol/L (\ie pH $= 12.89$ and $\beta = 10$). Chemical equilibrium in stress-free conditions is expected at $\beta = 1$, which implies $c_{Ca}=16.43$ mmol/L and $c_{OH}=32.86$ mmol/L (pH $= 12.51$) for the fixed $c_{OH}/c_{Ca}=2$ ratio assumed here to ensure charge neutrality of the solution.

For the dissolution reaction in \eqname\ref{eqCHstoi}, the activation energy $\Delta G^{o\ddag}_d$ and standard state concentration of the activated complex $c^{\ddag}_d$ are both computed from the rate constant of Ca(OH)$_2$ dissolution in \cite{bullard2007three}: $k = 0.66\;\mu$mol m$^{-2}$ s$^{-1}$. The conversion from $1\;\mu$mol m$^{-2}$ to number of reactions per $nm^2$, which are the simulation units, yields $c^{\ddag}_d = 10^{-6}N_a/10^{18} = 0.6022$ nm$^{-2}$. \eqname\ref{eqkconst} then provides $\Delta G^{o\ddag}_d = 74$ kJ/mol $= 122.9$ ag nm$^2$ ns$^{-2}$ in simulation units. For the precipitation reaction in \eqname\ref{eqCHstoi}, $c^{\ddag}_{pr} = c^{\ddag}_d$ is assumed, thus $\Delta G^{o\ddag}_{pr} = \Delta G^{o\ddag}_{d} +k_B T \ln(K_{eq,d}) = 45$ kJ/mol $=74$ ag nm$^2$ ns$^{-2}$. The activity coefficients of the activated complex are taken as $\gamma^\ddag_d = \gamma^\ddag_{pr} = 1$, as usual for reactions involving solids \citep{dominguez1998calculation}.

Deletion of existing particles and nucleation of trial particles are sampled using two different subcommunicators: $subA$ with 1 processor, and $subB$ with 2 processors. For particle nucleation, a cubic lattice of 50,656 trial particles is employed, with lattice spacing $r_{0,CH}/2$ in all three directions. The trial lattice is only created in a spherical shell region with inner and outer radii of 1.8 and 5 nm respectively, as shown in \figname\ref{figEx1init}; this avoids the computational cost of sampling unlikely precipitation events far from the surface of the initial crystal, while leaving room for enough growth or dissolution to reach a steady state regime with constant average reaction rate. For the trial particles to move into a local minimum of interaction energy, 600 steps of energy minimisation are performed using the aforementioned $quickmaske$ algorithm, with time step of 0.45 ps and a maximum displacement per step of $r_{0,CH}/200$. Such a short minimization is appropriate because, for the interactions used here, a particle only needs to move by $r_{0,CH}$ or less, to find its closest local energy minimum. The ions added/removed to/from the solution after particle dissolution/precipitation are distributed between the volume of solution $V_{sol}$ in the simulation box and the additional $\Delta V_{box}$ in proportion to their volumes. However, since $\Delta V_{box} \gg V_{sol}$, changes in ion concentrations and saturation index saturation $\beta$ within the simulation box are negligible here. After each accepted dissolution or precipitation event, a process of type $Every$ is invoked, which minimizes the total interaction energy by displacing all the real particles in the system; this is done \via 10,000 steps of the LAMMPS $quickmin$ algorithm, with timestep of 1 ps and a maximum displacement per step of 0.2 nm.

% ---------------------------
\subsection{Example 2: strain-rate effect on a crystal under stress-driven dissolution}\label{secEx2}
% ---------------------------

\begin{figure}[h]
\centerline{\includegraphics[width=0.95\textwidth] {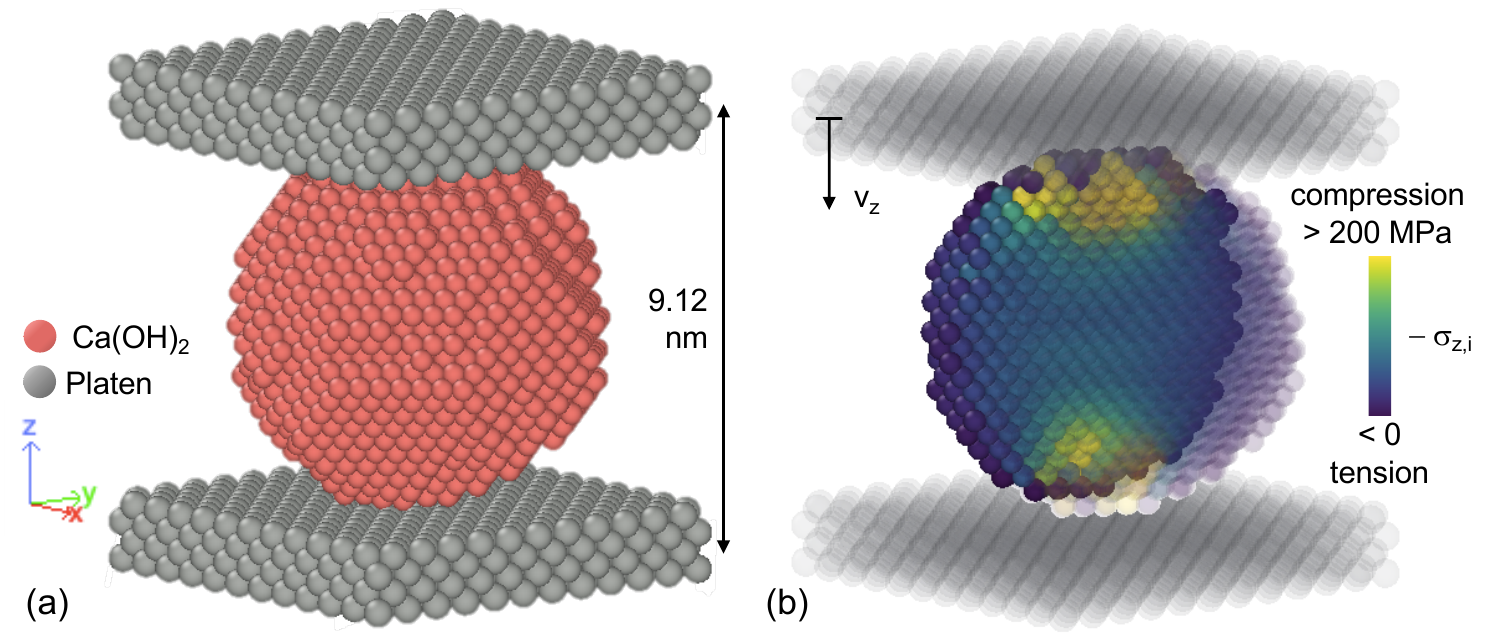}}
\caption{(a) Crystal of Ca(OH)$_2$ in initial contact with two platens; (b) Section of the crystal showing the distribution of per-particle virial stress $\sigma_{z,i}$ after a small downward displacement of the top platen.}
\label{figEx2init}
\end{figure}

The nanocystal of Ca(OH)$_2$ from the previous example is now brought into initial contact with two platens, \via a procedure detailed in \cite{masoero2023maske}: see \figname\ref{figEx2init}.a. The platens are arbitrarily discretized with particles of diameter $D_S = 0.6$ in an FCC lattice, hence with equilibrium distance $r_{0,S} = \sqrt[3]{\eta}D_{S} = 0.5427$. The mechanical interactions between particles in the platens are of the form in \eqname\ref{eqUij}, with elastic modulus typical of steel, $E_S = 200$ GPa, which yields a high interaction stiffness, $k_{0,S} = 208,397$ MPa$\cdot$nm  A high interfacial energy is then assumed between platen particles and the solution, $\gamma_S = 200$ mJ m$^-2$, which leads to $\varepsilon_{0,S} = 37.70$ MPa$\cdot$nm$^{3}$. The resulting cutoff for the platen particles is $r_{c,S} =  0.5617$ nm, \viz a strain at failure of \ca 3.5\%. The mechanical interactions between platen and Ca(OH)$_2$ particles are of the form in \eqname\ref{eqUijDiss}, with equilibrium distance $r_{0,CH-S} = 0.4856$ nm and interaction stiffness $k_{CH-S} = 33,200$ MPa$\cdot$nm, obtained from the already computed, individual $r_0$ and stiffness of the two particle types. The platen-Ca(OH)$_2$ bond energy is set to $\varepsilon_{CH-S} = 0$, making the interaction purely repulsive, \ie $r_{c,CH-S} = r_{0,CH-S}$ and no bond under tension.

The volume of the simulation box is set to $V_{box} = 8,000$ nm$^3$, with an additional volume in contact with it $\Delta V_{box} = 10,000$ nm$^3$. The initial volume of the crystal is given by the initial number of Ca(OH)$_2$ particles times their individual volume, $V_{0,CH} = N_{0,CH} \cdot V_{p,CH} = 4055 \cdot 0.0557 = 226$ nm$^3$, hence the initial volume of solution (in $V_{box}$ plus in $\Delta V_{box}$) is $V_{sol,tot} \approx 18,000$ nm$^3$. The change in concentration of Ca$^{2+}$ and OH$^-$ ions in solution caused by the dissolution of one Ca(OH)$_2$ can be approximately estimated by neglecting the changes in solution volume caused by the reaction (MASKE accounts for them, through the volume of the Ca(OH)$_2$ particle and the apparent volumes of solvated Ca$^{2+}$ and OH$^-$ ions given in \secname\ref{secEx1}). For Ca$^{2+}$, a dissolution event releases one molecule, hence changing $c_{Ca}$ approximately by $1/V_{sol,tot} \approx 92$ $\mu$mol/L; for OH$^-$ ions the change is $2/V_{sol,tot} \approx 184$ $\mu$mol/L. Therefore, in this example the dissolution of Ca(OH)$_2$ particles during a simulation may significantly change the concentrations of ions in solution, and thus the saturation index $\beta$.

Only the Ca(OH)$_2$ particles are allowed to dissolve and to nucleate, \via the same chemical reactions, thermodynamic and kinetic parameters, and trial precipitation lattice as in Example 1. The initial ion concentrations are set to $c_{Ca} = 0.01643$ mol/L and $c_{OH} = 0.03286$ mol/L, implying $\beta = 1$. This means that no sustained dissolution nor precipitation is expected in stress-free conditions. All the particles in both platens are constrained not to move under the action of forces; this is done using the $setforce$ fix in LAMMPS. However, the particles in the top platen are rigidly displaced towards the bottom platen by 0.01 nm (\via the $displace\_atoms$ command in LAMMPS) every $dt$ nanoseconds of simulated time in MASKE. This is done by using a continuous process of $mstoreLMP$ type in the MASKE input script. The chosen value of $dt$ controls the downward velocity of the top platen: $v_z = 0.01/dt$ nm/ns. The initial distance between the centres of mass of the two platens is 9.12 nm, hence the imposed strain rate is estimated in $\dot{\epsilon_z} \approx 0.01/dt/9.12 = 1.1\cdot 10^{-3}/dt$ ns$^{-1}$.

The platens put the Ca(OH)$_2$ crystal under compression: see \figname\ref{figEx2init}.b. The average compressive stress in the crystal, $\sigma_z$, is estimated as the per-particle virial stress in $z$ direction for each Ca(OH)$_2$ particles, $\sigma_{z,i}$ (computed in LAMMPS \via a $compute$ of type $stress/atom$), averaged over all the $N_{CH}$ particles of Ca(OH)$_2$ existing at any time in the simulation:
\begin{equation}
\sigma_{z,i} = \frac{1}{V_{p,i}} \sum_{j=1}^{N_{n,i}} \left( z_i\cdot F_{z,j}+ z_j\cdot F_{z,i} \right)
\label{eqSz}
\end{equation}
$V_{p,i}$ is the volume of the i$^{th}$ particle, j is the index of a neighbouring particle (of any type), $N_{n,i}$ is the number of neighbours mechanically interacting with particle $i$, $z$ is the Z coordinate of the two interacting particles, and $F_z$ is the Z component of the force on each particle stemming from the pairwise interaction. The presence of per-particle stresses entails a nonzero excess enthalpy in the rate equations for dissolution and precipitation. As a result, the compression is expected to trigger dissolution of Ca(OH)$_2$ even if the solution is initially at $\beta = 1$. As the compressive strain increases, dissolution will help relaxing the increasing $\sigma_z$, but the concurring increase in ion concentrations in solution, and thus of $\beta$, will try to slow down dissolution and even cause precipitation of new Ca(OH)$_2$ particles in stress-free regions of the crystal surface. All this will result in a complex kinetic balance between solution chemistry and mechanical stress, which will depend on the imposed strain rate and generate a non-trivial evolution of the crystal morphology through dissolution/precipitation events and plastic deformations.

% ---------------------------
% ---------------------------
\section{Results}
% ---------------------------
% ---------------------------

% ---------------------------
\subsection{Example 1: dissolution, growth, and equilibrium of a stress-free crystal}\label{secEx1Res}
% ---------------------------

This section considers the Ca(OH)$_2$ nanocrystal described in \secname\ref{secEx1}, with the aim of assessing: (1) whether the simulations correctly predict the transition from crystal dissolution to growth that, for stress-free crystals, is expected as the solution goes from undersaturated ($\beta <1$) to supersaturated ($\beta > 1$); (2) the variability in the results stemming from the random numbers involved in the Kinetic Monte Carlo algorithm (see \ref{secTimeApp} to appreciate the role of random numbers); (3) how the fineness of the lattice of trial particles may impact the results; (4) the pros and cons of using straight or net rates (\viz \eqnames\ref{eqr1d} and \ref{eqr1pr} rather than \eqnames\ref{eqrNd} and \ref{eqrNpr}).

\begin{figure}[h]
\centerline{\includegraphics[width=0.98\textwidth] {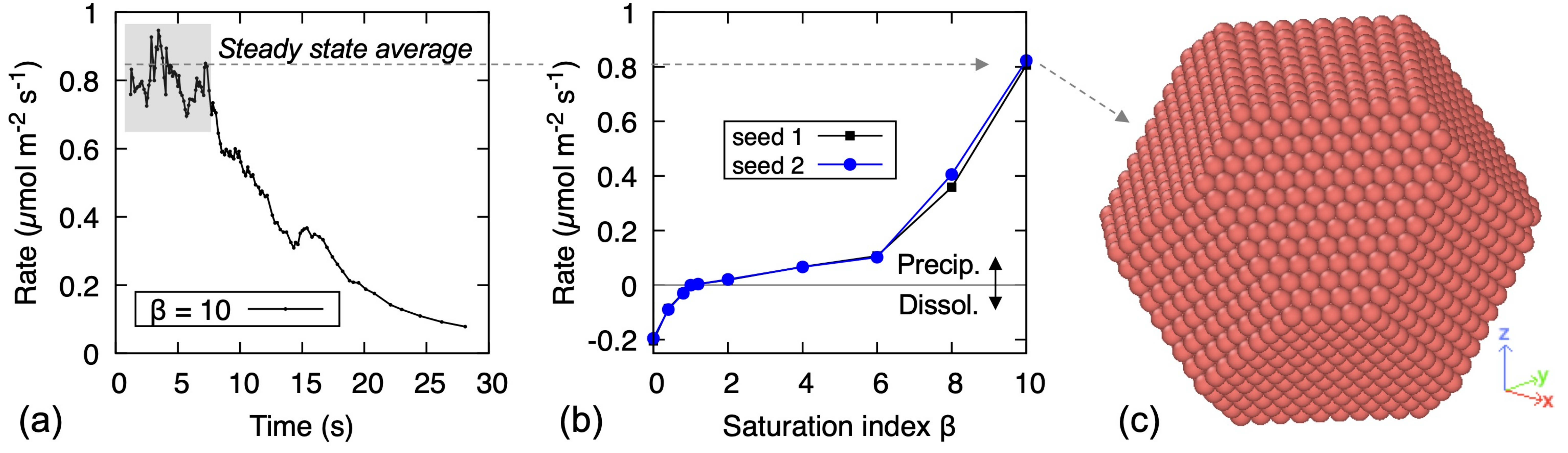}}
\caption{(a) Precipitation rate as a function of time for a Ca(OH)$_2$ crystal in a supersaturated solution with $\beta = 10$. The rate is normalized per unit area of crystal surface and per unit time; (b) Initial rates of dissolution and precipitation at various saturation levels, showing the expected transition across chemical equilibrium at $\beta = 1$ for two different sequences of pseudo-random numbers in the KMC algorithm (seed 1 and seed 2); (c) Final crystal configuration after precipitation at $\beta = 10$.}
\label{figEx1net}
\end{figure}

\figname\ref{figEx1net} shows precipitation and dissolution rates obtained from simulations using net rates. The rates are expressed per unit area of crystal surface $S$ and per unit time. To estimate $S$, knowing the volume $V_p$ of a Ca(OH)$_2$ particle and the number $N_{CH}$ of particles at any give time in a simulation, one can compute the total volume of the crystal $V_pN_{CH}$, and set $S=\pi(6V_pN_{CH}/\pi)^{2/3}$, having approximated the whole crystal as a sphere. \figname\ref{figEx1net}.a shows a typical curve of precipitation rate (averaged over the 100 KMC events before the current one) \vs simulated time. The time scale extends to various seconds, despite the nanometre scale of the system: this ability to sample long time scales irrespective of the length scale being considered, is a strength of the KMC approach. \figname\ref{figEx1net}.a also shows a steady state of constant rate in the initial part of the simulation; this happens for all $\beta \le 1$ and $\beta \ge 8$ values, but not when $1<\beta<8$. The steady state values are plotted against $\beta$ in \figname\ref{figEx1net}.b; for the $1<\beta<8$ range, the plotted rate is the one averaged over the first 100 KMC steps of a simulation. \figname\ref{figEx1net}.b shows that: (i) dissolution (\viz a negative precipitation rate) is correctly predicted when $\beta < 1$, as well as precipitation when $\beta < 1$ and and chemical equilibrium (\viz zero rate) when $\beta =1$. The figure also shows that the sequence of pseudo-random numbers used in the KMC algorithm has a negligible effect on the average rates (`seed 1' and `seed 2' indicate two difference sequences).

Transition State Theory (TST) predicts a linear relationship between rates per unit surface and $\beta$; by contrast the simulations, despite using rates that are based on TST, produce a nonlinearity in the $1<\beta<8$ range: see \figname\ref{figEx1net}.b. This effect is due to the crystal morphology. An overall linear relationship, as per TST, requires that kink sites are not exhausted during precipitation; this is possible if there is a very large number of such sites in the initial crystal, which is not the case for the nanocrystal simulated here. Alternatively, new kink sites must be created rapidly, which is possible if $\beta$ is large compared to the energy barriers of precipitating particles with fewer neighbours $n$ than a kink site, $\Delta G_n = \varepsilon_0 (n_{kink}-n)$, or if rare, energetically unfavourable fluctuations favour precipitation in under-coordinated sites, such as adatoms, which provide starting points for the growth of new layers. The latter scenario is impossible when net rates are used, as they average out all the energetically unfavourable fluctuations. As a result in \figname\ref{figEx1net}.b, when $1<\beta<8$, precipitation is short-lived as it only consumes the initial kink sites. The continuous exhaustion of kink sites causes a continuous decrease in rate, which explains why a steady state is not established in said range. When $\beta \ge 8$, instead, new kink sites can be formed on certain crystal facets and precipitation can proceed further and faster. However, when the layers growing from facets on which new kinks can form at $\beta = 10$ are exhausted, precipitation stops and the crystal ends up with the clear-cut morphology in \figname\ref{figEx1net}.c. Even higher $\beta$ values would enable formation of new kinks on other facets too, thus sustaining further precipitation; a systematic exploration of these $\beta$ thresholds is not of primary interest here, but an indication of this process at play will be given later in Example 2. A close inspection of \figname\ref{figEx1net}.b reveals a nonlinear increase in dissolution rate also as $\beta$ approaches 0. This happens because a very low $\beta$ enables the dissolution of sites that are more coordinated than kinks, creating vacancies (or `pits') on the crystal surfaces where new dissolution fronts can emanate, accelerating the overall process.

\begin{figure}[h]
\centerline{\includegraphics[width=0.98\textwidth] {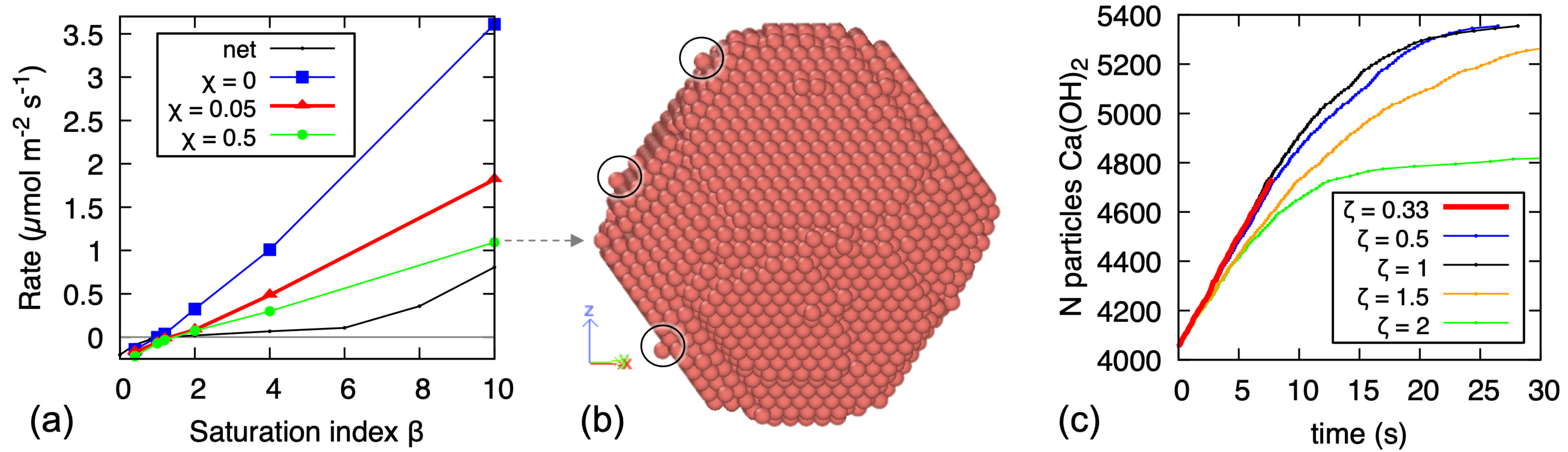}}
\caption{(a) Precipitation rate \vs $\beta$ curves from using straight rates; (b) a typical final morphology from simulations with straight rates and $\chi = 0.5$ (this one is for $\beta = 10$). A few adatoms are circled in black, which result from energetically unfavourable fluctuations and provide nucleation spots for the growth of new layers; (c) Effect of the fineness of the cubic trial lattice for precipitation, whose spacing is scaled here by a factor $\zeta$, such that the spacing is $\zeta\cdot r_{0,CH}/2$. The kinetics in (c) are for simulations with net rates, at $\beta = 10$.}
\label{figEx1latt}
\end{figure}

The effect of energetically unfavourable fluctuations is shown in \figname\ref{figEx1latt}.a,b, with simulation results from using straight rates. The $\chi$ parameter in the rate equations controls how prevalent the effects of fluctuations are. The explored values are $\chi = 0$ (as usual in Transition State Theory), which attributes all the excess enthalpy from mechanical interactions to the dissolution reaction only and maximizes the fluctuations, to $\chi = 0.5$, which equally distributes the excess enthalpy between dissolution and precipitation, minimizing fluctuations but not removing them (as opposed to net rates). \figname\ref{figEx1latt}.a shows that all the results from straight rates predict a linear relationship between overall rate and $\beta$, also in the $1\le \beta\le 8$ range where net rates were unable to nucleate new crystal layers. Straight rates instead always enable the formation and stabilization of adatoms (\eg those circled in \figname\ref{figEx1latt}.b) which allow for indefinite growth. A close inspection of \figname\ref{figEx1latt}.a reveals a slight shift of equilibrium (\ie rate = 0) towards $\beta$ values that are slightly greater than 1; this is the effect of relying on an estimated value of $V_t$ when explicitly sampling both dissolution and precipitation events (here $V_t$ was set to $V_{p,CH} \frac{r_{c,CH}}{r_{0,CH}}$).

All the simulations with straight rates, $\beta > 1$, and $\chi= 0.5$ produce the same morphology as in \figname\ref{figEx1latt}.b, which is reached when the entire nucleation lattice is covered by real particles; further growth is expected if bigger lattices were used. The same morphology is expected to emerge when $\chi = 0$ and $\chi = 0.05$, but in this case the simulations were stopped after a few thousands atoms precipitated, because of the high computational cost of sampling many unlikely fluctuations. Indeed, the precipitation of \ca 500 particles at $\beta = 10$ required \ca 10,000 KMC steps in simulations with $\chi = 0$ and 8,000 steps when $\chi = 0.05$, as opposed to 800 steps with $\chi = 0.5$, and 500 steps for net rates. On the other hand, \figname\ref{figEx1latt}.a shows that the precipitation rates compute with $\chi=0$ are significantly underestimated in all the other simulations with $\chi > 0$, as well as when net rates are used. This confirms that rare fluctuations are important for quantitatively capturing the nucleation of new layers. Therefore, when using MASKE one should judge on a case-by-case basis whether to use straight or net rates and, in case, which value of $\chi$, in order to balance the requirements of running rapid calculations, predicting qualitatively correct mechanisms, and obtaining quantitatively correct rates.

A last parametric study here concerns the impact of the fineness of the nucleation lattice on the resulting average rates. For simulations using net rates, \figname\ref{figEx1latt}.c shows the temporal increase in $N_{CH}$ at $\beta = 10$ when cubic nucleation lattices of different spacing $\zeta \cdot r_{0,CH}/2$ are employed. All simulations with lattice spacing less than $r_{0,CH}/2$, \viz with $\zeta \le 1$, return very similar results \footnote{The simulation with $\zeta = 0.33$ was stopped earlier because it entailed the sampling of \ca 1.4 million trial particles, as opposed to \ca 50,000 when $\zeta = 1$ and ca. 6,000 when $\zeta = 2$}. Coarser lattices however, with $\zeta > 1$, underestimate precipitation because they miss energetically convenient minima in the interaction energy landscape. This indicates that a nucleation lattice spacing of $r_{0,CH}/2$ or less is appropriate for the simulations and the mechanical interaction potentials considered here.

% ---------------------------
\subsection{Example 2: strain-rate effect on a crystal under stress-driven dissolution}\label{secEx2Res}
% ---------------------------

In this example, the Ca(OH)$_2$ crystal from the previous section is surrounded by a relatively small volume of solution, which is initially at chemical equilibrium with the crystal itself, \viz $\beta = 1$. The crystal is then subjected to a compressive strain with fixed rate, which should favour the dissolution of particles under high mechanical stress. The dissolving particles increase the concentration of Ca$^{2+}$ and OH$^-$ in solution, hence its saturation index $\beta$ too; this may cause precipitation of new Ca(OH)$_2$ particles in regions of the crystal surface that are under low mechanical stress. At high imposed strain rates, plastic deformations may occur too.

\begin{figure}[h]
\centerline{\includegraphics[width=0.85\textwidth] {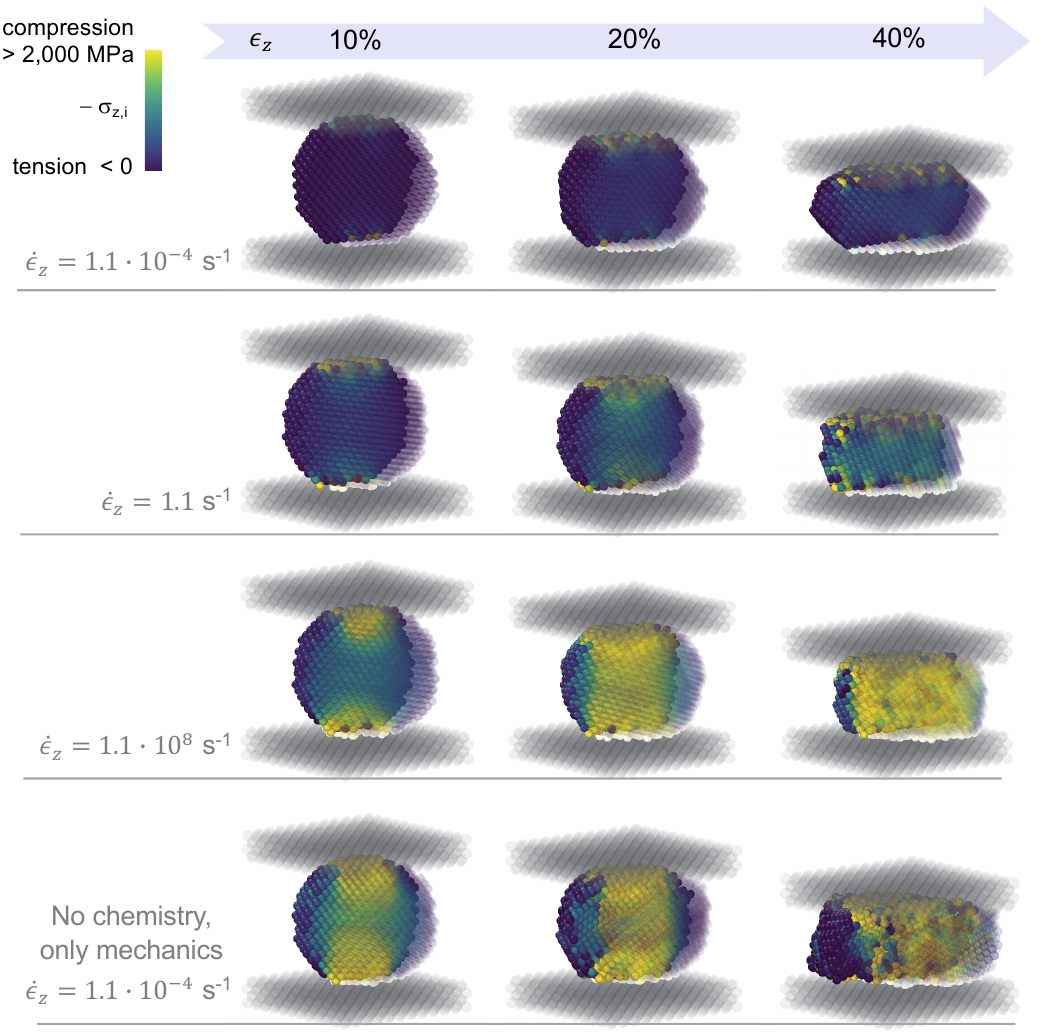}}
\caption{Snapshots from simulations of a Ca(OH)$_2$ nanocrystal deforming, dissolving, and reprecipitating under a compression imposed at a constant strain rate $\dot\epsilon_z$. The color map highlights the per-particle virial stress $\sigma_{z,i}$.}
\label{figEx2snaps}
\end{figure}

\figname\ref{figEx2snaps} shows the deformation and stress field of the Ca(OH)$_2$ crystal when compressed at different strain rates $\dot\epsilon_z$. The KMC algorithm in MASKE allows exploring a broad range of strain rates, here from $10^8$ s$^{-1}$ to $10^{-4}$ s$^{-1}$; this is valuable because the experiments typically impose rates in the order of the lowest values here, which are difficult to reach with other simulation techniques (\eg molecular dynamics; \cite{cao2019potential}). All the simulations in \figname\ref{figEx2snaps} employed net rates; some considerations on the effect of straight rates will be made later in this section.

At low strain rates, $\dot\epsilon_z \sim 10^{-4}$ s$^{-1}$ in \figname\ref{figEx2snaps}, the dissolution triggered by the per-particle stress (\via the excess enthalpy in the rate equations) is fast compared to the deformation process. As a result, the dissipation of local stress $\sigma_{z,i}$ through dissolution is efficient, therefore the local stress remains relatively low even at high $\epsilon_z$, as shown by the dark violet color in the $\epsilon_z = 40$\% snapshot in \figname\ref{figEx2snaps}. The same snapshot shows significant recrystallization, driven by the increase of $\beta$ in solution eventually triggering precipitation of new Ca(OH)$_2$ in low-stress regions.

Higher strain rates induce higher per-particle stresses $\sigma_{z,i}$: see $\dot\epsilon_z =1.1$ and $\sim 10^{8}$ s$^{-1}$ in \figname\ref{figEx2snaps}. For compatibility of displacements, the rate at which a layer of particles in contact with the platens dissolves must equal the velocity at which the top platen moves. A higher strain rate entails a higher platen velocity, which requires a higher dissolution rate, achieved by allowing a slightly higher strain before dissolution, which in turn increases the local stress and thus the excess enthalpy term in the rate. \figname\ref{figEx2snaps} also shows that little recrystallization takes place at high strain rates. This is because the precipitation rate scales linearly with $\beta$, and therefore with the number of dissolved particles (assuming that no reprecipitation takes places at all, which is a good approximation at high strain rates). At a given strain level $\epsilon_z$, the number of dissolved particles is basically the same irrespective of the imposed strain rate (for sufficiently high $\dot\epsilon_z$). By contrast, the strain rates have been increased by orders of magnitude here, as well as the dissolution rates required to accommodate the strain; in the resulting short timescales of the deformation/dissolution processes, precipitation events become rare. In the simulation with highest strain rate, $\sim 10^{8}$ s$^{-1}$, the animation of the full chemo-mechanical process in \figname\ref{figEx2snaps} shows that plastic shear slips along crystalline planes occur at high strain levels, $\epsilon_z \gtrsim 30\%$. Such slips are not recorded at lower strain rates, where the dissolution-induced relaxation of local stresses prevents the activation of plastic deformations.

\figname\ref{figEx2snaps} also shows results for a purely mechanical response, \ie with both dissolution and precipitation turned off. The simulation was conducted at a low strain rate $\dot\epsilon_z \sim 10^{-4}$ s$^{-1}$, but the same result would be obtained at higher strain rates because the particles are brought to static equilibrium between two successive execution of the platen-moving continuous process in MASKE (static equilibrium is imposed \via a MASKE process of type $Every$ performing energy minimization in MASKE). Therefore inertial effects do not contribute to rate-dependence here. The purely mechanical response in \figname\ref{figEx2snaps} is somewhat similar to the coupled chemo-mechanical one at high $\dot\epsilon_z \sim 10^{8}$ s$^{-1}$, but with some important differences too. At low strain levels (see $\epsilon_z = 10\%$ in \figname\ref{figEx2snaps}), even the limited dissolution taking place at $\dot\epsilon_z \sim 10^{8}$ s$^{-1}$ is sufficient to reduce the local stresses compared to the purely mechanical test. At $\epsilon_z \sim 20\%$ and $40\%$ the purely mechanical test displays a wider low-stress region near the surface of the crystal; the particles in this region do not contribute significantly to the overall mechanical response to the compression. Partial dissolution at $\dot\epsilon_z \sim 10^{8}$ instead moulds the crystal layers in contact with the platens, removing particles with high $\sigma_{z,i}$ that induce stress concentration, and thus enabling a more uniform mechanical response. Another distinctive aspect of the purely mechanical test is that extensive plastic deformations from shear slips already start at $\epsilon_z \approx 17\%$, and a fully developed central shear band controls the entire deformation response at $\epsilon_z \gtrsim 30\%$. In the dissolving crystal with $\dot\epsilon_z \sim 10^{8}$, instead, small and isolated, plastic shear slips only start appearing at $\epsilon_z \approx 30\%$.

\begin{figure}[h]
\centerline{\includegraphics[width=0.98\textwidth] {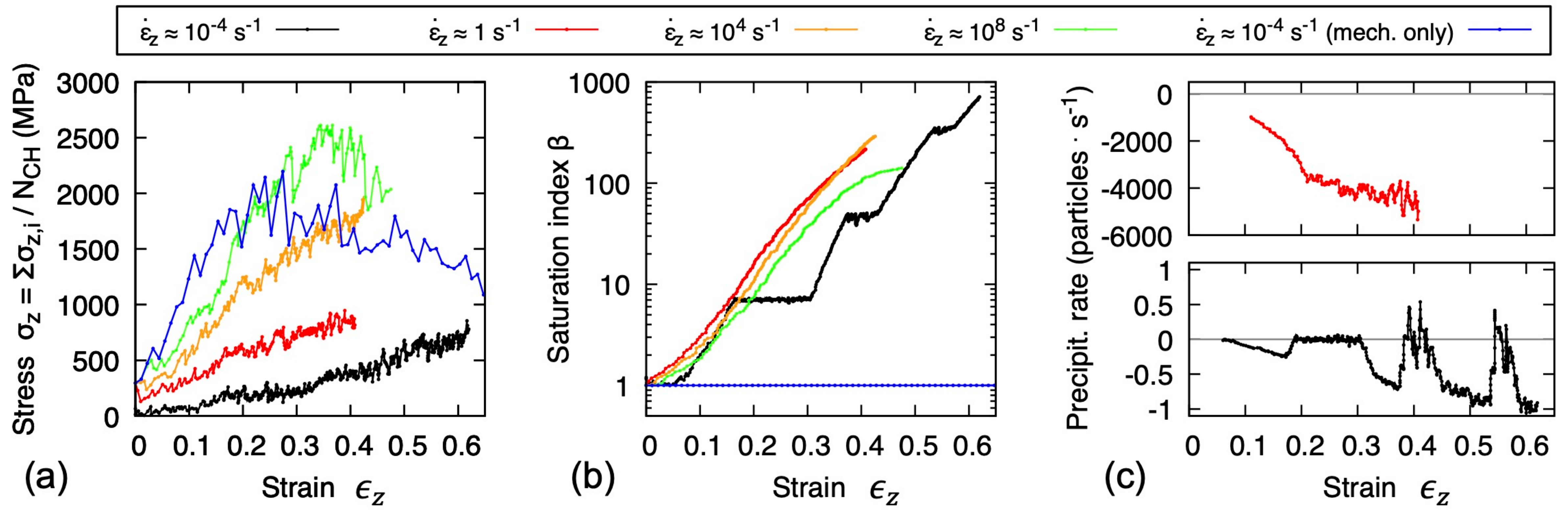}}
\caption{(a) Stress-strain curves for the Ca(OH)$_2$ nanocrystal from simulations at different strain rates $\dot\epsilon_z$; (b) Evolution of saturation index $\beta$ during the simulations; (c) Evolution of the precipitation rate (negative for dissolution), averaged over 100 KMC steps, from the simulations with $\dot\epsilon_z \approx 10^{-4}$ and $1$ s$^{-1}$.}
\label{figEx2graphs}
\end{figure}

\figname\ref{figEx2graphs}.a shows the stress strain curves computed at different strain rates $\dot\epsilon_z$. A higher $\dot\epsilon_z$ produces a higher average compressive stress $\sigma_z$; this reflects the already-discussed need for faster dissolution (whose rate depends on local stresses) to ensure compatibility of displacements at the platen-crystal interface. 

%how, at low strain rates, stress-driven particle dissolution helps relaxing the overall average stress $\sigma_z$ in the crystal. Indeed, at any given strain level $\epsilon_z$, the corresponding $\sigma_z$ in \figname\ref{figEx2graphs}.a increases with $\dot\epsilon_z$. There are then two interesting aspects of the stress-strain curves which will be discussed below: (i) the fact that the maximum stress in the simulation with  $\dot\epsilon_z \approx 10^8$ s$^{-1}$ is greater than that in the simulation with mechanics only, \ie with no dissolution nor precipitation allowed; (ii) the stepwise stress-strain curve at low strain rate $\dot\epsilon_z \approx 10^{-4}$ s$^{-1}$.
%

A counter-intuitive result in \figname\ref{figEx2graphs}.a is that crystals that are allowed to dissolve, when compressed at high rates, are stronger than crystals that are not allowed dissolve. This is indicated in \figname\ref{figEx2graphs}.a, by the higher peak $\sigma_z$ of the curve for $\dot\epsilon_z \approx 10^{8}$ s$^{-1}$, compared to the `mech. only' curve. The `mech.~only' curve lacks the mechanism of particle dissolution, which helps relaxing the local stresses on particles while ensuring compatibility of displacement at the crystal-platen interface. Such compatibility, in the `mech.~only' simulation, relies entirely on mechanical deformations; these are initially elastic and, when local stresses become too high, plastic shear slips are triggered and relax the local stress. One may thus expect that the `mech.~only' simulation, being more constrained, would produce a higher $\sigma_z$ than simulations where dissolution is possible too. This is indeed the case at low strain levels: see $\epsilon_z<0.2$ in \figname\ref{figEx2graphs}.a. However, the first extensive shear slips in the `mech.~only' simulation occur already at $\epsilon_z \approx 17\%$. After that point, subsequent shear slips effectively cap the peak stress $\sigma_z$ to \ca 2'000 MPa. By contrast, in the $\dot\epsilon_z \approx 10^{8}$ s$^{-1}$ simulation, dissolution of the most stressed particles limits the maximum local stress, which gets redistributed as a lower stress to the particles neighbouring the dissolved one. All this leads to a more uniform stress distribution in the crystal (see \figname\ref{figEx2snaps}) while also delaying the triggering of plastic slips until $\epsilon_z \approx 30\%$. As a result, the peak stress $\sigma_z$ in the  $\dot\epsilon_z \approx 10^{8}$ s$^{-1}$ is \ca 2'500 MPa, greater than in the `mech. only' case.

Another finding from \figname\ref{figEx2graphs}.a is the stepwise stress-strain curve at the lowest strain rate, $\dot\epsilon_z \approx 10^{-4}$ s$^{-1}$. Specifically, $\sigma_z$ increases with $\epsilon_z$ for a while, then it oscillates around a constant value while $\epsilon_z$ keeps increasing, until a new regime of increasing $\sigma_z$ is entered, and so forth. Three such steps can be counted in \figname\ref{figEx2graphs}.a. This behaviour is the result of the kinetic chemo-mechanical balance established during the simulations. At low rates $\dot\epsilon_z$, no plastic deformations are triggered and particle dissolution is the only mechanisms to accommodate the externally imposed $\dot\epsilon_z$. When compression starts, the first dissolution events release ions in solution, raising the saturation index $\beta$ for Ca(OH)$_2$ precipitation: see the increasing $\beta$ at $\epsilon_z < 0.15$ in the $\dot\epsilon_z \approx 10^{-4}$ s$^{-1}$ curve in \figname\ref{figEx2graphs}.b. With the net rate equations used here, Example 1 in the previous section has shown that no significant precipitation is possible when $\beta \lesssim 8$, hence $\beta$ keeps increasing. The increasing $\beta$ slows down dissolution but, since the overall rate $\dot\epsilon_z$ is imposed and must be matched, a compensating increase in dissolution rate is made possible through an increase in mechanical stress (which increases the excess enthalpy term in the dissolution rate equation). This explains the regimes of increasing $\sigma_z$ in \figname\ref{figEx2graphs}.a, which indeed correspond to regimes of increasing $\beta$ in \figname\ref{figEx2graphs}.b. At $\beta \approx 8$, particle precipitation consumes any further ion that dissolution releases in solution, so that the average rate (precipitation minus dissolution over 100 KMC steps) oscillates around zero: see the curve for $\dot\epsilon_z \approx 10^{-4}$ s$^{-1}$ in \figname\ref{figEx2graphs}.c when $0.15 < \epsilon_z < 0.3$. As a result, $\beta$ now remains constant while $\epsilon_z$ increases, and the constant overall rate $\dot\epsilon_z$ implies that the stress $\sigma_z$ must remain constant too. This explains the matching plateaus in \fignames\ref{figEx2graphs}.a and \ref{figEx2graphs}.b. The first plateau continues until exhausting all the surface sites where precipitation at $\beta \approx 8$ is possible: \cf the final morphology in \figname\ref{figEx1net}.b. Then a new regime of increasing $\beta$ and $\sigma_z$ is entered, until a new plateau occurs at $\beta\approx 50$, when precipitation on other exposed crystal surfaces becomes possible too. The new plateau again corresponds to a net dissolution rate oscillating around zero, in \figname\ref{figEx2graphs}.c. All this describes a complex, kinetic equilibrium between composition of the solution, stress state in the crystal, and microstructural evolution.

The stepwise shape of the stress-strain curve is lost at higher strain rates, as shown by the $\dot\epsilon_z \approx 1$ s$^{-1}$ curves in \figname\ref{figEx2graphs}. Additional results omitted here show a progressive reduction and disappearance of the plateaus for intermediate rate values between $\dot\epsilon_z \approx 10^{-4}$ s$^{-1}$ and 1 s$^{-1}$. This means that, despite some particles do still precipitate when $\beta \gtrsim 8$, they cannot compete any more with the faster rate of the dissolving particles. Consistently, when $\dot\epsilon_z \approx 1$ s$^{-1}$ the saturation index $\beta$ increases monotonically in \figname\ref{figEx2graphs}.b and the average dissolution rate never goes to zero in \figname\ref{figEx2graphs}.c. The $\beta(\epsilon_z)$ curves in \figname\ref{figEx2graphs}.b for $\dot\epsilon_z \approx 1$ and $10^{+4}$ s$^{-1}$ are similar, meaning that dissolution progresses at a similar rate. This is expected because, in a scenario where precipitation is negligible, the total amount of dissolved ions depends only on the accommodated strain $\epsilon_z$. If $\beta$ is approximately identical for $\dot\epsilon_z \approx 1$ and $10^{+4}$ s$^{-1}$, then $\sigma_z$ is necessarily greater at the higher imposed rate, since only $\sigma_z$ now can speed up the dissolution rate to match $\dot\epsilon_z$. Less intuitively, the $\beta(\epsilon_z)$ curve for $\dot\epsilon_z \approx 10^8$ s$^{-1}$ in \figname\ref{figEx2graphs}.b is initially similar to those for $\dot\epsilon_z \approx 1$ and $10^{+4}$ s$^{-1}$, but it significantly diverges from them  at $\epsilon_z \approx 0.2- 0.3$, leading to less concentrated solutions at high strain levels $\epsilon_z$. The reason for this lies in the triggering of plastic deformations at high rate, $\dot\epsilon_z \approx 10^8$ s$^{-1}$, which cap the stress level (see $\sigma_z$ at $\epsilon_z \gtrsim 0.3$ in \figname\ref{figEx2graphs}.a) and provide another mechanism to accommodate the imposed strain. With a capped $\sigma_z$, the increasing $\beta$ reduces the dissolution rate, which in turn reduces the slope of the $\beta(\epsilon_z)$ curve in \figname\ref{figEx2graphs}.b. The resulting interdependence between plastic deformations and composition of the solution is another non-trivial result of the chemo-mechanical coupling in MASKE.

All the results in this section are from simulations where the $allpar$ mechanism in MASKE employed net rates of dissolution and precipitation, as per \eqnames\ref{eqrNd} and \ref{eqrNpr}. Additional tests were conducted using straight rates too, as per \eqnames\ref{eqr1d} and \ref{eqr1pr}, with $\chi$ values between 0 and 0.5. At high strain strain rates, $\dot\epsilon_z \gtrsim 1$ s$^{-1}$, straight and net rates produce almost identical results; this is expected, since the main effect of using straight straight is to favour precipitation, which plays a negligible role at high $\dot\epsilon_z$. Instead, at low $\dot\epsilon_z \lesssim 1$, simulations using straight rates produced lower $\beta(\epsilon_z)$ than those from net rates at same $\dot\epsilon_z$. This agrees with the findings from Example 1 in the previous section, where straight rates were shown to enable precipitation at lower $\beta$ values. Therefore, simulations with straight rates are indeed expected to produce plateaus in $\beta$ and $\sigma_z$ at lower $\epsilon_z$ than simulations with net rates. These aspect, however, should be studied further because the efficiency loss from using straight rates significantly limited the maximum $\epsilon_z$ that could be reached during the simulations. For the purpose of this paper, the results presented here for net rates have already demonstrated how MASKE can predict the emergence of a complex kinetic balance between chemical and mechanical processes over long time scales.

%%*****************************************
\section{Conclusion and outlook}
%%*****************************************

The article has introduced MASKE, a kinetic simulator of microstructural evolution driven by chemical reactions and mechanical stress. Solid domains are modelled as mechanically interacting particles, while the surrounding environment, which was an aqueous solution in the examples here, is modelled implicitly through the concentrations of molecular species in it. The composition of the solution co-evolves with the solid phases, in respect of mass conservation, all while dissolution and precipitation reactions transform molecules from the solid into solvated species and vice versa.

The uniqueness of MASKE stems from several key strengths:
\begin{enumerate}
\item MASKE implements an off-lattice, rejection-free, Kinetic Monte Carlo (KMC) framework to simulate the dissolution and precipitation of particles. This framework gives access to realistic time scales that are relevant for the experiments;
\item MASKE features original chemo-mechanical rate equations for dissolution and precipitation, which capture the effect of mechanical stress on the reaction rates. To simulate the deformations that induce mechanical stress, the particles are free to move off-lattice, but this implies that infinite positions are to be sampled for possible particle precipitation. To treat this issue, a discretization method has been proposed which approximates the integral of the precipitation rate over the whole simulation box by considering only a finite number of trial particles;
\item The reaction rates are implemented in both straight and net forms. Depending on the application, the user can opt for the computational efficiency of net rates, or for more computationally demanding but also more precise (mechanistically and quantitatively) simulations using straight rates;
\item The KMC sampling of discrete dissolution/precipitation events is coupled with the explicit integration of continuous processes. In the article, this coupling has been exploited to impose a strain rate. Other continuous processes are currently being implemented, in particular one that simulates the metabolic activity of bacteria, explicitly modelled as particles as in \cite{li2019nufeb}, for the purpose of modelling biomineralization in self-healing concrete \cite{bagga2022advancements};

\item In MASKE, the user can define bespoke chemical reactions and mechanisms in a chemistry database file. MASKE is also interfaced with the LAMMPS library, through which the user can specify complex initial microstructures, loading conditions, and interaction potentials. As a result, a wide variety of problems with disparate chemistries can be simulated;
\item MASKE features a two level parallelisation, ready for use in High Performance Computing. The simulations in this article used both levels of parallelisation, but on simple examples that required only 3 processors in total (despite one simulation already featured 1.4 million trial particles). Applications of MASKE involving more processors have been presented elsewhere \citep{coopamootoo2020simulations,alex2023carbonation}.
\end{enumerate}

A number of additional features provide promising avenues for further development. These in particular may involve:
\begin{enumerate}[\hspace{0.5cm}a. ]
\item Coarse-graining the reaction rates to reach microstructural scales. The $allpar$ mechanism presented here is only accurate for unimolecular reactions, but \cite{shvab2017precipitation} already proposed coarse-grained rate expression for nanoparticles with diameter of \ca 10 nm. The version of MASKE that is currently on GitHub already includes a mechanism for particles with diameter of $\sim 1$ $\mu$m, which will be presented in a separate article;
\item Allowing particles to partially grow or dissolve, whereas now they can only appear or disappear at once. This would be desirable when using coarser particles, to avoid unrealistically large, sudden changes in stress (and thus in stress-dependent reaction rates) when a full particle is deleted or added. Partial dissolution and growth would thus be important for capturing well certain phenomena, such as crystallization pressure \citep{masoero2023maske}. Even better if the partial growth or dissolution of particles were anisotropic. Such anisotropy would at least require ellipsoidal particles, which would be rather straightforward to include in MASKE, since ellipsoidal particles and appropriate potentials for them already exist in LAMMPS \citep{berardi1998gay}. However, partial growth and dissolution (either as discrete KMC events or as continuous processes) will require additional implementation and a new, probably less efficient, way of sampling energy changes $\Delta U$ to compute the excess enthalpy terms in the reaction rates;
\item Improving the model of the solution, allowing for heterogenous local concentrations of solvated molecules, which may also react chemically within the implicit fluid. For the reactions in solution, there is ongoing work to couple MASKE with the PHREEQC library \citep{parkhurst1999user}. Heterogeneous concentrations would be desirable for simulations at larger length scales, and the resulting concentration gradients would prompt diffusion that can be implemented in MASKE as a continuous process to be integrated in time.
\end{enumerate}

Two examples in this article have shown that MASKE, already in its current form, can capture  interesting coupled phenomena that challenge other simulation methods. Example 1 confirmed that MASKE correctly predicts the dissolution, growthm and chemical equilibrium of a nanocrystal of Ca(OH)$_2$, depending on the concentrations of Ca$^{2+}$ and OH$^-$ ions in the surrounding solution. Straight rate equations have been shown to effectively allow the nucleation of new crystal layers through energetically unfavourable fluctuations. Example 2 then addressed the effects of chemo-mechanical coupling, by imposing a compressive strain rate to the crystal. The results showed that a complex kinetic balance is established between dissolution rates, chemical composition of the solution surrounding the crystal, elastic stress state in the crystal, and plastic deformations too. A family of curves is obtained, which quantifies the stress-strain response of the crystal as a function of the imposed strain rate. These curves offer a constitutive models for simulations al larger scales, \eg continuum simulations based on the Finite Element Method. All in all, MASKE offers a new way to reconstruct the chemo-mechanical kinetics of evolving microstructures. This capability can be leveraged to simulate the formation and behaviour of new materials in computer-aided design, and to predict the degradation of materials for both conventional and accidental scenarios of future exposure.

\bigskip
%\textbf{Acknowledgments}

%The research leading to this publication benefitted from EPSRC funding under grant No. EP/R010161/1 and from support from the UKCRIC Coordination Node, EPSRC grant number EP/R017727/1, which funds UKCRIC’s ongoing coordination.

%E.M. acknowledges the support of the TU1404 COST Action, EU Framework Programme Horizon 2020. E.M. also thanks the Matua Campus of Politecnico di Milano, for supporting his stay at Politecnico di Milano in April 2017. The work of G.C. was supported under NRC grant NRC-HQ-60-14-G-0003.

%\clearpage
%\section*{References}

\bibliographystyle{elsarticle-harv} 
\bibliography{../../bibliocement}

%\bibliographystyle{chicaco}
%\bibliography{diluzio}

\newpage

\appendix

\section{Details on system description}\label{secAppA}

The main article has presented a typical simulation in MASKE as featuring explicit particles, typically representing solid units, and an implicit fluid surrounding them: see \figname\ref{figSystem}. The current version of MASKE assumes prismatic boxes (\textit{orthogonal} in LAMMPS syntax) but extensions to other box shapes, \eg triclinic ones, are within scope for future implementation. MASKE accepts all boundary conditions in LAMMPS, but for now only fixed and periodic conditions have been tested. LAMMPS assigns a \textit{type} to each particle, which MASKE uses to associate possible events to sets of particles: for example, MASKE may associate a certain reaction formula to the dissolution of type 1 particles, and a different formula to particles of type 2. One would often associate different types to different solid species, but MASKE does not require that, as it does not track the chemical composition nor the type of individual particles; the user is responsible for attributing events and reactions consistently with particle types. Such implementation is appropriate when a particle represents a homogeneous, mono-phase solid, such as a nanocrystal; future versions of MASKE may include tracking of the internal structure of particles, \eg to simulate partial growth or dissolution of particles each representing a polycrystal.

Particle shapes are also defined in LAMMPS, \via the \textit{atom\_style} command. For example, the \textit{sphere} style includes information on per-particle radius, whereas \textit{ellipsoid} includes per-particle shape through radii along three principal axes. MASKE for now assumes spherical particles, hence only uses per-particle radii. Future implementation may also include other atom\_styles from LAMMPS, \eg to consider individual atoms and directly apply MASKE at the atomistic level, or to employ ellipsoidal particles for simulating richer coarse-grained morphologies.

The particles interact \via effective potentials, which are defined and managed in LAMMPS: \eg see \figname\ref{figSystem}. The particle displacements caused by the interactions are also computed by LAMMPS, with typical methods such as energy minimization or molecular dynamics. Currently MASKE extracts per-particle interaction energies from LAMMPS, obtained through a \textit{pe/atom} compute. MASKE uses these energies to quantify the change in total energy of the system, $\Delta U_i$, generated by adding or deleting the generic particle $i$. This $\Delta U_i$ informs the reaction rates. For pairwise interactions, $\Delta U_i$ is simply twice the per-particle energy of particle $i$ read from LAMMPS. Currently MASKE assumes that this relationship holds, hence that pairwise interactions are used in LAMMPS (these include user-defined, tabulated, pair potentials). Higher-order interactions, such as three-body or else, will need future implementation and a different, probably more time-consuming, approach to compute $\Delta U_i$ in MASKE. 

The implicit solution surrounding the solid particles is defined and managed directly in MASKE, which tracks the number of molecules and molar concentrations  for various, user-defined, solvated species. Molarities require the user to specify the molecular volume of each solvated species, so that the total volume of the solution can be tracked while chemical reactions alter it. Future versions of MASKE may adopt molalities instead. The concentrations of solvated species are assumed to be uniform in the simulation box, which is the same as assuming infinitely fast diffusion. This hypothesis is realistic at small scales, \eg below the micrometre. Diffusion algorithms may be included in future implementations. Currently, MASKE does not minimize the free energy of the solution through speciation, neither it considers the kinetics of chemical reactions between molecules in solution, unless these reactions produce or dissolve solid particles as a result. There is however no obstacle to implementing these features and, indeed, current development is underway to couple MASKE with PHREEQC to include speciation in solution.

During a typical simulation, molecules disappear from solution to form new solid particles, or vice versa. During these transformations, MASKE preserves the total number of molecules, as long as the user has consistently associated particle types to solid species, and to chemical reactions with correct stoichiometry. The volume of molecules in solution typically differs from that of the same molecules when combined in a solid; as a result, the total volume of solid particles plus implicit solution may change as the system evolves. The MASKE variable $voidV$ tracks the difference between the volume of the simulation box, $V_{box}$, and the volume of solid plus solution in it. If $voidV<0$, then the total volume of the system has expanded beyond $V_{box}$. Presently, $voidV$ is just recorded for information, but in the future it may be used to change the volume of the box or to estimate the pressure of the fluid.

A last element of the model is an additional volume of solution, $\Delta V$, implicitly attached to the simulation box. $\Delta V$ initially features the same molecules with same concentrations as in the solution in $V_{box}$. Currently there is no process in MASKE to exchange molecules between $V_{box}$ and $\Delta V$, but dissolution/precipitation events taking place in $V_{box}$ can use molecules from $\Delta V$ if allowed by the user.

\section{Processes and time line in MASKE}\label{secTimeApp}

MASKE allows for both discrete and continuous processes to take place contextually. An example of discrete process may be the precipitation of a new solid particle, which occurs instantaneously after a time $\Delta t$ from a previous event, \eg precipitation or dissolution of another particle. A continuous process, instead, advances gradually with time; examples may be the diffusion of molecules in solution, displacements being applied to certain particles with a given rate, or solid particles growing or dissolving only in part. Continuous processes are typically governed by differential equations that MASKE integrates over time, with a user-specified time step $dt_i$ for the $i^{th}$ process. The user also provides a number of integration steps $n_i$, so that whenever the $i^{th}$ continuous process is invoked, MASKE integrates it over a time period $\Delta t_i = dt_i \cdot n_i$ (future implementations may include adaptive $dt_i$ and criteria to stop integrating processes that have reached equilibrium). 

\begin{figure}[h]
\centerline{\includegraphics[width=0.45\textwidth] {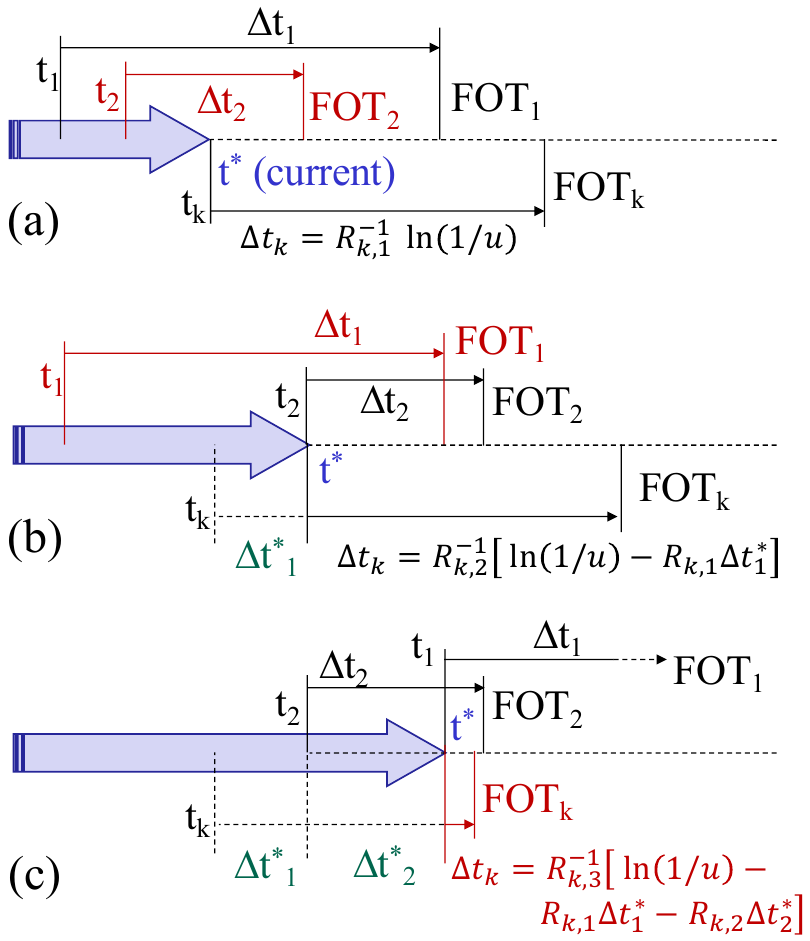}}
\caption{Time line in MASKE with continuous processes 1 and 2, and with the discrete KMC processes grouped under the letter $k$. The subfigures show three successive steps in MASKE, where the process to be carried out (in red) is the one with nearest future time of occurrence, FOT. (a) carries out continuous process 2, (b) carries out continuous process 1, and (c) carries out a discrete event. In (a) the current time $t^*$ was set by a discrete event that had just been carried out, hence the cumulative rate of discrete events $R_{k,j}$ was the first one to be calculated since accepting a discrete event, \ie $n_{cp}=0$ in \eqname\ref{eqDtkTdep}. Subfigure (c) eventually brings back a scenario where a discrete event is accepted, hence $n_{cp}$ is reset after that and the following step will be analogous to the one in (a). Note how $\Delta t_i$ and $FOT_i$ of continuous events remain constant (in the current implementation of MASKE), whereas $\Delta t_k$ and $FOT_k$ generally change whenever $t^*$ advances.}
\label{figTimeline}
\end{figure}

To stagger discrete and continuous processes, MASKE employs the concept of future time of occurrence, $FOT$. For each $i^{th}$ continuous process, $FOT_i = t_i + \Delta t_i$, where $t_i$ is the last time that the process has reached through previous integration: see processes 1 and 2 in \figname\ref{figTimeline}.a.
The $FOT_k$ of discrete events, instead, is determined by a single $\Delta t_{k}$ that accounts for all discrete events, in the spirit of rejection-free Kinetic Monte Carlo (KMC). Consider, for example, a simulation box containing $N$ solid particles and two, user-defined, possible discrete processes: dissolution of any of said $N$ particles, and nucleation of one new particle at $M$ possible locations; this means $N+M$ possible discrete events. MASKE computes a rate for each of these events, and cumulates them into a total KMC rate $R_{k}$. The time increment associated to one discrete event taking place is thus:
\begin{equation}
\Delta t_{k} = R_{k}^{-1} \ln\left(\frac{1}{u}\right)  \label{eqDtkSimple}
\end{equation}
where $u$ is a random number extracted with uniform probability in (0,1]. \eqname\ref{eqDtkSimple} is only correct if the rates of all the discrete events remain constant between two successive realizations of a discrete events; generally this is not the case in MASKE, since continuous processes taking place meanwhile might change the composition of the solution or the stress state, both of which can impact the rate of individual events. To account for this, a more general expression for $\Delta t_{k}$ will be given shortly below.

\figname\ref{figTimeline} depicts how $FOT$s are used to advance a simulation in MASKE. Without losing generality, assume that just before \figname\ref{figTimeline}.a, a discrete event had taken place, determining the value of the current time in MASKE to be $t^* = t_k$. MASKE computes the $\Delta t_i$ and $FOT_i$ of all continuous processes ($i=1,2$ in the figure, with last occurrence times $t_1$ and $t_2$ respectively), and the $\Delta t_k$ and $FOT_k$ of a generic discrete event, using \eqname\ref{eqDtkSimple}. In \figname\ref{figTimeline}.a, the continuous process 2 has the smallest $FOT$ and is therefore carried out, setting the new $t^*$ and $t_2$ in \figname\ref{figTimeline}.b to coincide with $FOT_2$ from \figname\ref{figTimeline}.a. MASKE computes the new $FOT_2$ in \figname\ref{figTimeline}.b, but it must also update $FOT_k$ to account for the change in rates that the execution of continuous process 2 might have caused. To this end MASKE employs a time-dependent formulation of KMC, whereby the remaining $\Delta t_k$ follows the relation:
\begin{equation}
\sum_{j=1}^{n_{cp}}{\Delta t^*_j R_{k,j}} +  \Delta t_k R_{k,n_{cp}+1} = \ln\left(\frac{1}{u}\right) \label{eqDtkTdep}
\end{equation}
$n_{cp}$ is the number of time increments ($\Delta t^*_j$) that continuous processes have brought about since the last discrete event. For the scenario in \figname\ref{figTimeline}.b, $n_{cp}=1$ due to the continuous process that prevailed in \figname\ref{figTimeline}.a, and the cumulative rate of all discrete events changed from $R_{k,1}$ in \figname\ref{figTimeline}.a to $R_{k,2}$ in the current state. Therefore \eqname\ref{eqDtkTdep} yields $\Delta t_k = R_{k,2}^{-1}\left[\ln\left(\frac{1}{u}\right) - \Delta t^*_1 R_{k,1}\right]$ as written in \figname\ref{figTimeline}.b. In some cases \eqname\ref{eqDtkTdep} may return a negative $\Delta t_k$; when this happens MASKE issues a warning and approximates $\Delta t_k = 0$, thus ensuring that a discrete event will be carried out next.

Going back to \figname\ref{figTimeline}.b, $FOT_1$ is now the smallest hence process 1 is carried out leading to a new $t^*$ and $t_1$ in \figname\ref{figTimeline}.c. In \figname\ref{figTimeline}.c, $FOT_1$ and $FOT_k$ are updated and the latter happens to be the smallest one, hence a discrete event is carried. The individual discrete event to execute (\eg removing or inserting one specific particle) is selected out of the list of all possible discrete events contributing to $FOT_k$ with a probability that, for event $i$ out of the $M+N$ possible ones in our example, is:
\begin{equation}
P_i 	\propto \sum_{j=1}^{n_{cp}}{\Delta t^*_j r_{k,i,j}} +  \Delta t_k r_{k,n_{cp}+1} \label{eqPi}
\end{equation}
$r_{k,i,j}$ is the rate of event $i$ when the $j^{th}$ continuous process was executed since the last discrete event took place (\ie since $t^*$ in \figname\ref{figTimeline}.a). $r_{k,i,n_{cp}+1}$ is the latest rate of the same event, which contributed to the cumulative $R_{k,n_{cp}+1}$ in \eqname\ref{eqDtkTdep} and that now gets multiplied times the $\Delta t_k$ from \eqname\ref{eqDtkTdep}.

The execution of a discrete event in \figname\ref{figTimeline}.c brings back an analogous scenario as in \figname\ref{figTimeline}.a. This, however, does not mean that the order in which processes are carried out will remain the same: for example, after \figname\ref{figTimeline}.c, process 2 will be executed twice before calling again process 1.

%*************************************
\section{Code structure and input script}\label{secStructApp}
%*************************************

\begin{figure}[h]
\centerline{\includegraphics[width=0.49\textwidth] {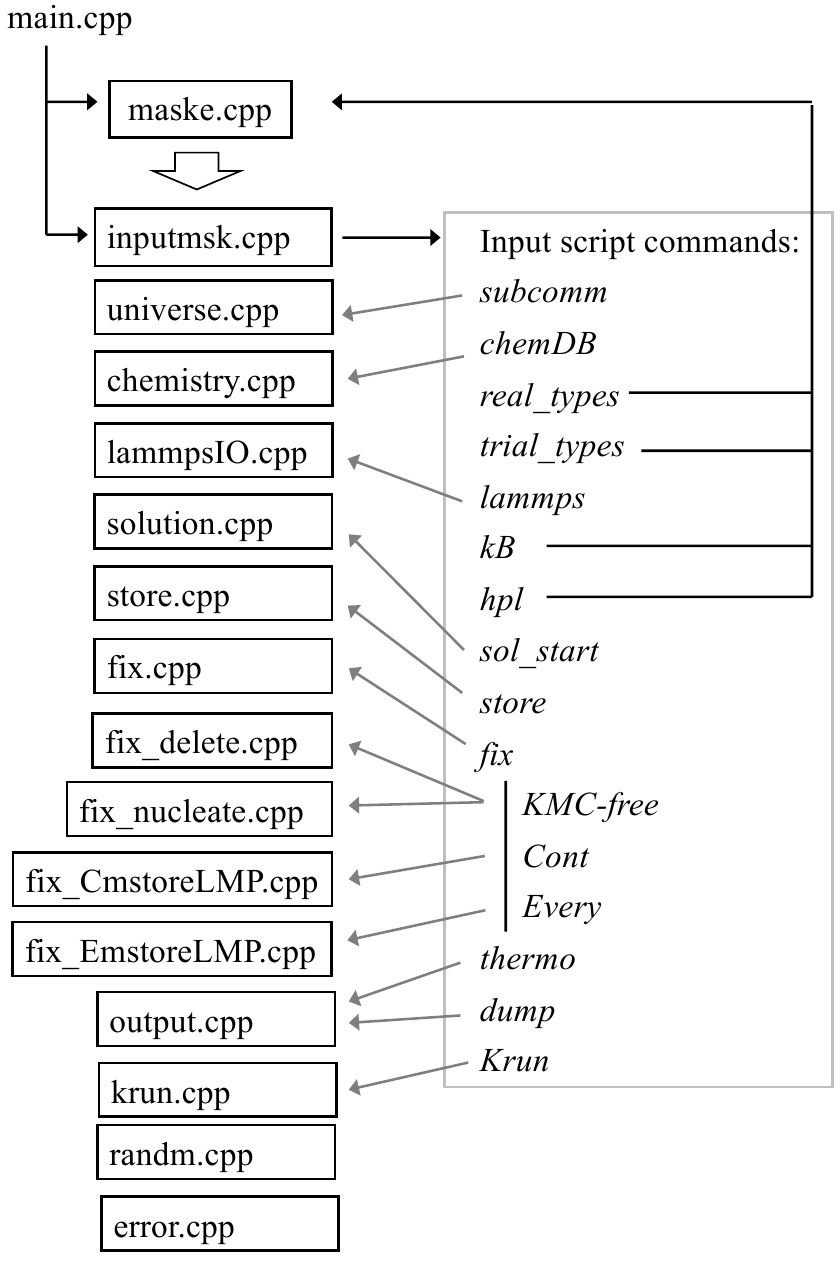}}
\caption{Scheme of the classes forming MASKE and their main relationships to user commands in the input script.}
\label{figStructure}
\end{figure}

\figname\ref{figStructure} shows the code structure of the MASKE software. MASKE features two levels of parallelization using the MPI library. The user decides how many processors to involve, $n_p$ and assigns them to $n_{sc}$ different sub-communicators. This assignment is carried out by the \textit{MPI\_Commm\_Split} command in the \textit{universe} class. Each sub-communicator may be responsible for carrying out a subset of commands and for sampling a subset of all the discrete events and compute their individual rates $r$. This is important because sampling discrete events is usually the most time-consuming task in MASKE. Furthermore, each sub-communicator may perform its tasks and sampling using multiple processors in parallel. For example, the user may use 12 processors in total for a simulation, assigning 8 to sub-communicator $subA$ and 4 to $subB$. $subA$ may sample the deletion of $N$ particles, which entails computing per-particle interaction energy using LAMMPS; for this operation, LAMMPS can operate in parallel using the 8 processors in $subA$. Meanwhile $subB$ may run a separate instance of LAMMPS using 4 processors to sample the creation of new particles at $M$ possible locations.

The main class in the code is \textit{maske}, which creates and eventually deletes all the other objects and which hosts several pointers to output files, flags, and parameters that are useful across the code: see \figname\ref{figStructure}. Another fundamental class is in \textit{pointers.h}, where MASKE employs the same strategy as in LAMMPS to allow any object to directly include and access any other object. Several classes produce support objects that are used throughout the code: \textit{error} logs error messages and stops a simulation when an error occurs; \textit{lammpsIO} creates LAMMPS instances and calls LAMMPS operations \via one-line strings like those in a LAMMPS input script; \textit{output} creates MASKE output files (more details later) and features functions to write individual entries into them; \textit{randm} features functions to generate pseudo-random sequences of numbers (based on user provided seeds) and to extract numbers from such sequences when needed (\eg $u$ in \eqname\ref{eqDtkTdep}).

The flow of user-defined operations is dictated by an input script, \ie a text file that the object \textit{inputmsk} reads and executes line by line. The two only exceptions to the line by line execution are the \textit{loop} and the \textit{store multi} commands. The \textit{loop} command makes MASKE repeat multiple times a number of input lines: its working is not detailed here for brevity. The \textit{store multi} command will be explained later. 

The \textit{subcomm} command is to create the desired number of sub-communicators and, for each of them, specify a name, assign a number of processors, create a LAMMPS instance if needed, and provide a seed for a sequence of pseudo-random numbers. 

The \textit{chemDB} command loads a text file with all the information required to describe the chemistry of the system and its chemical kinetics. This information includes data on individual species in solution or forming the solids, thermodynamic and kinetic parameters such as interfacial energies, activation energies, stoichiometric coefficients and equilibrium constants of possible chemical reactions, and mechanism according to which chemical reactions are combined together in the rate equations. A later dedicated section will provide more details on the structure and content of the file loaded by \textit{chemDB}. The information gathered by \textit{chemDB} is stored in the \textit{chemistry} object.

The \textit{real\_types} and \textit{trial\_types} commands are for the user to list the LAMMPS particle types associated to real and trial particles. The latter are used for sampling the possible formation of new particles at various locations in the simulation box. The vectors of real and trial particle types are stored in the \textit{maske} object.

The \textit{lammps} command passes a string with LAMMPS syntax directly to LAMMPS for direct execution. This is a very important command to set up a simulation box, to decide the \textit{atom\_style} and units in LAMMPS, to create particles if needed or to import a configuration from a previous LAMMPS dump file, to set up the interaction potentials, to run any preliminary energy minimization, to set up LAMMPS variables, computes and output files, etc. The $lammps$ command allows specifying which sub-communicator should execute the line; if the keyword \textit{all} is used, all the processors with an active LAMMPS instance will execute the line using LAMMPS. A type of LAMMPS command that are known not to work through single-line execution are those related to loops, hence a bespoke \textit{loop} command has been implemented in MASKE, as already mentioned.

The \textit{kB} and \textit{hpl} commands specify the values of the Boltzmann and Planck constants in the units chosen by the user; these units should be consistent with those set in LAMMPS. The values of the constants are recorded in the \textit{maske} object.

The \textit{sol\_start} command initializes the implicit solution in MASKE by listing the names and corresponding molar concentrations of the solvated species, as well as the of the solvent species. The names listed here must correspond to species that have been defined in the chemistry file previously read by the \textit{chemDB} command. \textit{sol\_start} also provides the temperature of the solution in Kelvin degrees and the $A$ and $B$ constants of the solvent as per Debye-H\"{u}ckel theory, in consistent units (\eg $A = 0.51$ and $B = 3.29$ nm$^{-1}$ for water). The command also assigns the initial volume of void space in the simulation box, the additional volume $\Delta V$ of implicit solution, the volume of voids in $\Delta V$, and a unit conversion factor to convert the user-decided units of volume into liters (\eg 10$^{-24}$ if the user is working in nm). Having read these inputs, \textit{sol\_start} calls the \textit{solution} object to compute the initial volume occupied by the solution in the simulation box and in $\Delta V$, respectively $SVol$ and $dVSVol$. The former is obtained as $SVol = V_{box} - voidV - SolidV$, where the volume of the simulation box $V_{box}$ and the volume occupied by solid particles $SolidV$ are both extracted from LAMMPS, assuming spherical particles. This implies that, in the current version of MASKE, the user should not change the simulation box size nor insert, delete, or alter the radii of particles after using the \textit{sol\_start} command, or MASKE will return an inconsistent $SVol$. Once $SVol$ is computed, the \textit{solution} object employs the user-provided initial concentrations to compute the number of molecules in solution for each solvated species in the box and then, knowing the apparent volume of each solvated species including the solvent, the number of solvent molecules is obtained to match $SVol$ itself. An analogous process is followed to compute $dVSVol$ in $\Delta V$ and the number of solvated molecules for each species there, except that there are no solid particles in $\Delta V$.

The \textit{store} command stores information for later, repeated use. Five styles of storing are currently implemented: four of them store a one-line string each and are used by the \textit{fix\_nucleate} object, described later, when sampling the insertion of new particles. These four styles are: \textit{region} and \textit{lattice}, which stores the LAMMPS commands to create respectively a region and a lattice, \textit{DV} which stores the LAMMPS command to create a variable quantifying the volume of a cell in the lattice, and \textit{minimize} which stores parameters to perform energy minimization in LAMMPS using a new and bespoke minimizer that MASKE adds as a user-defined package to LAMMPS when compiling the code. The package is called \textit{USER-MASKE} and can be found in the \textit{src} folder of the LAMMPS compiled as a submodule of MASKE. This minimizer, called \textit{quickmaske}, is analogous to \textit{quickmin} in LAMMPS but it treats each particle separately when applying and damping their displacements. This overcomes a limitation of all the minimizers in LAMMPS, which perform poorly when some particles occupy positions with very high energy, \eg largely overlapping with other particles. This is a frequent scenario when sampling insertion of many new particles at random places and running a single energy minimization for all of them, as the \textit{fix\_nucleate} object does in the current version of the code (more details later). The fifth and last \textit{store} style is \textit{multi}, which stores multiple strings for objects such as \textit{fix\_CmstoreLMP} and \textit{fix\_EmstoreLMP}, described later.

The \textit{fix} command sets the various processes determining the evolution of the system in MASKE. Three types of fixes are currently implemented: \textit{KMC-free} which sample and execute discrete events using the rejection-free KMC approach described above, \textit{Cont} which integrate continuous processes, and \textit{Every} which are invoked after $n_s$ processes, either discrete or continuous, have been carried out ($n_s$ is input by the user). Depending on the type of fix, MASKE expects a different set of parameters, which the user must provide in the input script and that MASKE records in the \textit{fix} object. Two parameters that are common to all fixes are the name assigned to the fix and the name of the sub-communicator that is responsible for sampling and executing it, when needed. \textit{fix Every} is the only type of fix which can, and often should, be run by all the sub-communicators via the \textit{all} keyword.

Two types of \textit{fix KMC-free} are currently implemented in MASKE: \textit{delete} and \textit{nucleate}, which respectively sample and execute the dissolution and precipitation of solid particles through the corresponding objects \textit{fix\_delete} and \textit{fix\_nucleate}. These objects are crucial in MASKE, hence they will be described in detail in later sections. For both of them, the user must specify: (i) the LAMMPS type of real particles that the fix would delete or create if an event of theirs is executed, (ii) a chemical mechanism from those defined in the chemistry file loaded by \textit{chemDB}, to compute the rates of individual events; (iii) a \textit{sol\_out} option specifying whether the molecules released into the solution, or removed from it, after executing an event will be released or removed only in the simulation box (\via the \textit{box} option) or also in the additional volume $\Delta V$ in proportion to the $SVol/dVSVol$ ratio (\textit{box+dV} option). In addition to these, for the \textit{nucleate} type, the user must also indicate the names of a stored region, lattice, and lattice cell size, as well as the LAMMPS type of trial particles for sampling precipitation, and the diameter of the particles that may precipitate (under the current assumption in MASKE of spherical particles).

Only one type of \textit{fix Cont} and one type of \textit{fix Every} are currently implemented, both of which simply run a sequence of LAMMPS commands previously stored using a \textit{store multi} command. The LAMMPS commands are executed respectively by the \textit{fix\_CmstoreLMP} and \textit{fix\_EmstoreLMP} objects when appropriate.

The \textit{thermo} command creates a text file to print output data every $n$ steps in MASKE, \ie one entry is recorded 
whenever $n$ events (discrete or continuous) have been carried out ($n$ is user-defined, as is the file name). The thermo file records $m$ user-chosen quantities, organized in a table with $m+2$ columns. The first two columns always record the MASKE step number and the MASKE time $t^*$ corresponding to the quantities in the current entry, \ie in the rest of the row being recorded. The quantities in the other $m$ columns may be of three types: \textit{conc\_name} prints the concentration in solution of a molecule defined in the \textit{chemDB} and called \textit{name} (for example, \textit{name} may be \textit{H2O} or \textit{CaCO3}); \textit{lmp\_v\_name} prints the content of a scalar variable previously defined in LAMMPS and called \textit{name} (for example, the user might have defined a variable called $press$ in LAMMPS, monitoring the pressure in the box); \textit{lmp\_c\_name} is analogous to \textit{lmp\_v\_name} but it prints the scalar resulting from a compute in LAMMPS instead of a variable.

The \textit{dump} command is the same as the \textit{dump} command in LAMMPS, in that it creates a file recording the configuration of the system (\ie particle positions and types) as well as various user-decided per-particle quantities such as  radii, stress per atom, etc. The only difference here is that the user inputs the number of MASKE steps between two successive records of the configuration, as opposed to the \textit{dump} command in LAMMPS which prints the configuration every so many steps in LAMMPS. Both the \textit{thermo} and \textit{dump} commands are carried out by the \textit{output} object.

The \textit{Krun} command invokes the \textit{krun} object to launch a simulation, performing MASKE steps until covering a user-specified period of time. This means that the user does not decide how many steps MASKE will perform, but rather by how much the current \textit{Krun} will advance $t^*$.

\section{Chemical database}\label{secChemDBApp}
The previous section mentioned that the \textit{chemDB} command loads a file containing data on the chemistry of the system. This file is loaded line by line, each line starting with a specific keyword.

The \textit{molecule} keyword creates species that can be used to initiate the solution (see the \textit{sol\_start} command in the previous section) or to define chemical reactions later in the \textit{chemDB} file. The keyword is followed by a user-defined name of the molecule, some quantities that are useful if the molecule is later  associated to a solid phase (namely quantities that describe the size and shape of the molecule in a solid) and other quantities that are useful for molecules that are later associated to the solution: apparent volume in solution, hydrated radius, ionic charge, and the parameter $b$ used in the Debye-H\"{u}ckel theory and discussed in the main body of this article.

The \textit{gammax} keyword is to record a value for the activity coefficient of an activated complex in standard state, $\gamma^\ddag$. The \textit{DGx} keyword is to record the standard state activation free energy of an activated complex, $\Delta G^\ddag$, as well as the concentration of said complex $c^\ddag$, and the dimensionality of the concentration: 2 if the  per unit area, 3 if per unit volume. All quantities recorded \via \textit{gammax} and \textit{DGx} can later be associated to certain chemical reactions, and will inform the rate equations from Transition State Theory described in the main body of the article.

The \textit{reax} keyword defines a chemical reaction, assigning a user-defined name to it. The attribute \textit{simple} indicates that a chemical reaction is just one-step (a \textit{chain} attribute can also be used, to combine a series of simple reactions, but this will not be discussed here for brevity). Take for example the simple reaction of calcium hydroxide dissolution:
\begin{equation}
\mathrm{Ca(OH)_{2}(s) \rightarrow Ca^{2+} (aq) + 2OH^- (aq) } \label{eqStoiApp}
\end{equation}
The \textit{reax} keyword imports previously recorded values of $\gamma^\ddag$, $\Delta G^\ddag$, and $c^\ddag$ to associate to this reaction, as well as an input value for the equilibrium constant $K_{eq}$. The stoichiometry in \eqname\ref{eqStoiApp} is then recorded by assigning the molecule Ca(OH)$_2$ to the solid (called `foreground' in MASKE and assigned \via the $fgd$ attribute) with a stoichiometric coefficient -1, because the reaction deletes one such molecule, and assigning the molecules Ca$^{2+}$ and OH$^-$ to the solution (`background' in MASKE, \via the $bkg$ attribute) with stoichiometric coefficients +1 and +2 respectively. An additional parameter is also needed: a real number $\chi$ between 0 and 1 used in the rate equations to distribute the excess free energy from mechanical stress between forward and backward reactions.

The \textit{sen} keyword records values of solid-solution interfacial energies that will later be associated to reaction mechanisms. Such mechanisms are defined using the \textit{mech} keyword; they are the last set of entries needed in the \textit{chemDB} file and they inform the \textit{fix} commands of \textit{KMC-free} type in the main input script, as discussed in the previous section. A \textit{mech} keyword requires a user-defined name for the mechanism and the name of a previously-defined chemical reaction to attach to it. The user then specifies the type of mechanism, which is important to determine how a number of chemical reactions of specified time would assemble to generate (or dissolve) the target particle. For now, only one type of mechanism is implemented in MASKE, called \textit{allpar}. This mechanism assumes that all the reactions required to delete or create a particle take place in parallel, hence the time to dissolve or form a new particle equals the time for a single unimolecular reaction to take place. This is exact only if the volume of a particle is the same as the volume of the individual molecules involved in a single reaction, \eg the volume of one Ca(OH)$_2$ solid molecule in the example above. The \textit{allpar} mechanism also assumes that the reaction rates follow the modified version of Transition State Theory presented in the main body of this article. The last attribute to associate to a mechanism is whether, in calculating the rates, it should consider straight or net reactions, also discussed in the main body of the article.

%************************
\section{Krun}\label{secAppkrun}
%************************

The main loop in MASKE, invoked by the \textit{Krun} command, advances the timescale through a combination of discrete KMC events and stepwise integration of continuous processes. As an example, consider a system made of three particle types, $A$, $B$, and $C$,  whose deletion or nucleation are sampled as discrete, \textit{KMC-free} events. Consider also that a continuous process (\textit{Cont} type) is taking place, \eg some particles are displaced with a given rate to apply strain. Finally, consider that after any process is executed, the interaction energy of all real particles in the system is minimized through a process of type \textit{Every}. Let us then assume that two subcommunicators are employed, $subA$ and $subB$, each running two processors, for a total of 4 processors being employed in the simulation. $subA$ is tasked to sample events for particles of types $A$ and $C$, namely the deletion of $N_A$ and $N_C$ real particles and the nucleation of new particles of type $A$ only at $M_A$ possible locations; let us call these events $fix\_delA$, $fix\_delC$, and $fix\_nucA$. Similarly, $subB$ samples $N_B$ deletions and $M_B$ nucleation sites for particles of type $B$, through events $fix\_delB$ and $fix\_nucB$. Processes of \textit{Cont} and \textit{Every} type are also tasked to one subcommunicator (or to all of them) but now it is not important to specify which one(s) since MASKE eventually synchronizes all processors after a process is executed.

\begin{figure}[h]
\centerline{\includegraphics[width=0.49\textwidth] {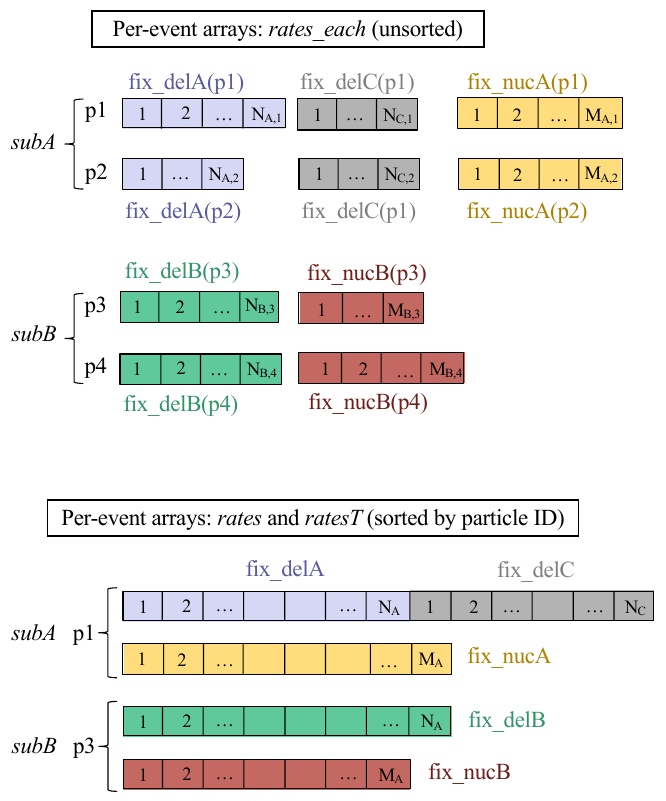}}
\caption{Rate arrays for four discrete \textit{KMC-free} processes ($fix\_delA$, $fix\_delB$, $fix\_nucA$, and $fix\_nucB$) distributed between two subcommunicators, $subA$ and $subB$, and four processors, $p1-p4$. Processors $p1$ and $p3$ are the submasters of $subA$ and $subB$; they gather and sort by particle ID all the rates coming from processor-specific $rate\_each$ arrays.}
\label{figProcs}
\end{figure}

\figname\ref{figKrunFlow}, in the main body of the article, shows a flow chart for the $Krun$ loop. The first step is for the $fix\_nucleate$ object to create trial particles at $M$ locations for each discrete \textit{KMC-free} event of type $nucleate$ being considered. In our example, $subA$ and $subB$ respectively create $M_A$ and $M_B$ trial particles at various locations. The creation of such particles is managed by the LAMMPS interface, which enables parallelization \via domain decomposition; this means that, since each subcommunicator features two processors in our example, each processor will manage a subset of the trial particles, \ie processors $p1$ and $p_2$ forming $subA$ will respectively create and manage $M_{A,1}$ and $M_{A,2}$ trial particles with $M_{A,1} + M_{A,2} = M_A$, and analogously for $p_3$ and $p_4$ in $subB$ with its $M_B$ trial particles.
%
%\begin{figure}[h]
%\centerline{\includegraphics[width=0.99\textwidth] {Figures/Fig5_krunFlow.pdf}}
%\caption{Flow chart of the main kinetic loop.}
%\label{figKrunFlowApp}
%\end{figure}
%
The $Krun$ loop then proceeds with sampling all the possible, discrete, \textit{KMC-free} events to compute their rates. This is done by the $fix\_nucleate$ and $fix\_delete$ objects, which perform operations that depend on user-specified mechanisms and involve one particle per event. For example, to inform the rate of a $delete$ event involving particle $i$ of type $A$ the processor hosting it in $subA$ may compute the change in interaction energy associated to removing that particle. In this way, each processor computes the rates $r_{k,i}$ of the events that are associated to the particles that LAMMPS hosts in that processor itself; \figname\ref{figProcs} helps clarifying this, with $p1$ for example computing rates for the deletion of $N_{A,1}$ and $N_{C,1}$ particles and the nucleation of the $M_{A,1}$ trial particles in it. Each processors stores these rates into two $rates\_each$ arrays, one in $fix\_delete$ and the other one in $fix\_nucleate$, as also depicted in \figname\ref{figProcs}. The $rates\_each$ arrays in each subcommunicator are then assembled by the submaster processor in it, leading to two arrays per subcommunicator containing respectively the $rates$ of all $delete$ and $nucleate$ events sampled there: see \figname\ref{figProcs}. The $rates$ arrays are sorted in increasing order of particle ID, these latter being defined in LAMMPS and extracted from there.

The cumulative of the $rates$ arrays across all subcommunicators is the $R_{k,n_{cp+1}}$ term in \eqname\ref{eqDtkTdep}, which for the first step in the loop coincides with $R_k$ in \eqname\ref{eqDtkSimple} hence suffices to compute $\Delta t_k$ and thus the $FOT_k$ of any one discrete event. The $Krun$ loop then samples the $\Delta t_i$ and thus $FOT_i$ of all continuous processes, which simply consists in multiplying together the user-defined integration time step $dt_i$ and number of integration steps $n_i$ of each $Cont$ process $i$. The overall simulation time is thus advanced to $t^* = \min\limits_{k,i} (FOT)$, \ie the smallest FOT out of $FOT_k$ for a discrete event and $FOT_i$ of all $i$ continuous processes. 
The process with smallest FOT is then carried out. If the process is of $Cont$ type, its execution simply entails the integration over $n_i$ steps with operations specified in an object that depends on the type of continuous process, \eg $fix\_CmstoreLMP$ in \figname\ref{figStructure}. The execution of a continuous process is carried out either by one or by all subcommunicators; for example, processes of type $fix\_CmstoreLMP$ are always executed by all subcommunicators. Instead, when a process is executed only by one subcommunicator, MASKE synchronizes all the other processors to recover the same system before progressing with the loop. 

When the chosen process is instead of \textit{KMC-free}, discrete type, the $krun$ object first updates the $ratesT$ arrays. These arrays are similar to the $rates$ array, in that each subcommunicator has one stored for $delete$ events and one stored for $nucleate$ events in its submaster processor: see \figname\ref{figProcs}. However, unlike the $rates$ arrays, the generic $i^{th}$ entry in the $ratesT$ arrays is obtained by taking its current value and adding the product between $r_{k,i}$ and $\Delta t_k$, where $r_{k,i}$ is the corresponding rate value stored in the $rates$ array. In other words, the $ratesT$ arrays store the $\sum_{j=1}^{n_{cp}}{\Delta t^*_j r_{k,i,j}}$ terms in \eqname\ref{eqPi}, and now the latest contributions to them, $\Delta t_k r_{k,n_{cp}+1}$ in \eqname\ref{eqPi}, is being added. As a result, $krun$ can now select one discrete event to carry out with probability $P_i$ as per \eqname\ref{eqPi}.

The execution of a discrete event entails either the deletion or the creation of a particle, in the current version of MASKE. This is carried out by all subcommunicators at the same time. Following the stoichiometry specified in the $chemDB$ file, the discrete event adds or removes molecules from the implicit solution (see the $reax$ keyword in \secname\ref{secChemDB}, and specifically the $bkg$ attribute in it). Therefore the composition of the implicit solution is modified accordingly, recomputing the new concentrations of all the solvated molecules based on the new number of molecules in solution and on their corresponding apparent volumes, which change the total volume of the solution itself. Also this operation is carried out by all subcommunicators. After having executed a discrete, \textit{KMC-free} event, the $ratesT$ arrays are zeroed across all subcommunicators, to restart the sums in \eqnames\ref{eqDtkTdep} and \ref{eqPi} (note that the term $\sum_{j=1}^{n_{cp}}{\Delta t^*_j R_{k,j}}$ in \eqname\ref{eqDtkTdep} is just the cumulative of all $ratesT$ arrays).

After any process is executed, be it continuous or discrete, the trial particles associated to the various nucleation events (of \textit{KMC-free} type) are brought back to their initial positions. Then processes of type $Every$ are run, provided that the correct, user-specified number of loops have been completed since their last execution.  Outputs are then printed out every so many (user-specified) number of loops; the outputs include MASKE's $thermo$ and $dump$ files described in \ref{secStructApp}. This triggers the beginning of the next loop, and thus the next step in MASKE. The loop ends when the simulated time $t^*$ is greater than the end time specified by the user \via the input command $Krun$: see\ref{secStructApp}.

%*****************************************
\section{Sampling deletion and nucleation rates}\label{secSampleApp}
%*****************************************

In the current version of MASKE, the sampling of discrete \textit{KMC-free} events is limited to the possibility of deleting or creating (nucleating) a full particle. When prescribing a deletion or nucleation process in the main input script of MASKE, the user specifies a mechanism in the $ChemDB$ file to attach to the process, which in turn invokes one of the user-specified chemical reactions. A reaction involves a number of molecules in solution (called 'background', in the $bkg$ attribute of the reaction) and a number of molecules in the solid phase (called 'foreground', \via the $fgd$ attribute). The number of reactions to nucleate a generic particle $j$ is the closest integer to:
\begin{equation}
n_{r,j,nuc} =  \frac{V_{p,j}}{\sum_i \nu{_i}v_{m,i,fgd}}\label{eqnmd}
\end{equation}
where $V_{p,j}$ is the volume of particle $j$ and $v_{m,i,fgd}$ is the volume of the $i^{th}$ foreground molecule involved in the reaction, with $\nu_i$ its stoichiometric coefficient. For particle dissolution, the expression is the same as in \eqname\ref{eqnmd}, except $-\nu_i$ should replace $\nu_i$.

For now, MASKE only features one mechanism, called $allpar$, whereby the nucleation or deletion of an entire particle, irrespective of its size, is fully determined by one single chemical reaction, assumed to take place at the same time for all the solid molecules in the particle. This means that the time to fully dissolve a particle, $\Delta t_{p,j}$ equals the time for one single chemical reaction to take place, $r^{-1}_1$, where $r_1$ is the rate of that chemical reaction. This mechanism is correct when a particle is only as large as the sum of solid molecules involved in one reaction, \ie $V_{p,j}=\sum\limits_i \nu{_i}v_{m,i,fgd}$ leading to  $n_{r,j,nuc} = 1$ in \eqname\ref{eqnmd}.  For larger particles, the other extreme case would be to consider all reactions taking place in series, in which case the particle nucleation or dissolution time would be $\Delta t_{p,j} = \sum\limits_{s=1}^{n_{r,j}} r_s^{-1}$, where $r_s$ is the rate of the $s^{th}$ reaction in the series. More realistic mechanisms for larger particles would involve a combination of reactions taking place in series and in parallel, such as in classical and non-classical nucleation and growth/dissolution theories. The interested reader can find an example of such a mechanism in \cite{shvab2017precipitation}.

The $allpar$ mechanism in MASKE uses a modified version of Transition State Theory (TST) for the rates of single dissolution/nucleation reactions, leading to the unit rate expressions in \eqnames\ref{eqr1d} and \ref{eqr1pr} in the main body of this article. There it is pointed out how the $c^{o\ddag}$ and $\Delta G^{o\ddag}$ terms (standard concentration and activation free energy of the activated complex) are related to the reaction rate constant $k$:
\begin{equation}
k = \kappa \frac{k_BT}{h}\frac{c^{o\ddag}}{\gamma^\ddag_d} \exp\left( \frac{-\Delta G^{o\ddag}}{k_BT} \right) \label{eqkconstApp}
\end{equation}
\eqname\ref{eqkconstApp} indicates that the arbitrary choice of $c^{o\ddag}$ has an effect on the value of $\Delta G^{o\ddag}$ to be employed. Consider a set of experiments of dissolution or precipitation conducted at different temperatures $T$. The Gibbs free energy can be expressed from its enthalpy $H$ and entropy $S$ terms: $\Delta G^{o\ddag} = \Delta H^{o\ddag} - T \Delta S^{o\ddag}$. One can thus rearrange \eqname\ref{eqkconstApp} to obtain:
\begin{equation}
\ln \left( \frac{k}{c^{0\ddag}} \cdot \frac{h \gamma^\ddag}{\kappa k_B T} \right) = \frac{\Delta S^{o\ddag}}{k_B} - \frac{\Delta H^{o\ddag}}{k_B} \frac{1}{T} \label{eqkexp}
\end{equation}
If the $\exp\left( \frac{-\Delta H^{o\ddag}}{k_BT} \right)$ term is assumed to lead the temperature dependence, a set of experimental data obtained at different $T$ would yield a linear plot of $\ln \left( \frac{k}{c^{0\ddag}} \cdot \frac{h \gamma^\ddag}{\kappa k_B T} \right)$ as a function of $T^{-1}$, with $\frac{\Delta S^{o\ddag}}{k_B}$ as the intercept when $T^{-1} \rightarrow 0$ and $\frac{\Delta H^{o\ddag}}{k_B}$ as the slope. Therefore, for a given choice of $c^{o\ddag}$, the linear plot provides the corresponding $\Delta G^{o\ddag}$. If $c^{o\ddag}_{new}$ is a new, different choice of $c^{o\ddag}$, the numerical value of $k$ expressed in this new units will change to $k_{new}$ (note from \eqname\ref{eqkconst} that the dimensionality of $c^{o\ddag}$ determines that of $k$; for example, if $c^{o\ddag}$ is 1 mol/m$^2$ then $k$ will be in mol/m$^2$ too). This will lead to a new linear plot yielding different values of enthalpy and entropy, \ie:
\begin{equation}
\ln \left( \frac{k_{new}}{c^{0\ddag}_{new}} \cdot \frac{h \gamma^\ddag}{\kappa k_B T} \right) = \frac{\Delta S^{o\ddag}_{new}}{k_B} - \frac{\Delta H^{o\ddag}_{new}}{k_B} \frac{1}{T} \label{eqkexpnew}
\end{equation}
By writing $\frac{k_{new}}{c^{0\ddag}_{new}} = \frac{k_{new}}{c^{0\ddag}} \cdot \frac{c^{0\ddag}}{c^{0\ddag}_{new}}$, one finds that the following relationships hold: $\Delta H^{o\ddag}_{new} = \Delta H^{o\ddag}$ and $\Delta S^{o\ddag}_{new}= \Delta S^{o\ddag} + k_B \ln \left (\frac{c^{0\ddag}}{c^{0\ddag}_{new}} \right)$. All this shows that the user must be careful to provide, as inputs, consistent values for $c^{o\ddag}$ and $\Delta G^{o\ddag}$. The conversion of $\Delta G^{o\ddag}$ values from its original $c^{o\ddag}$ to a new $c^{0\ddag}_{new}$ requires an additional entropy term due to the difference in standard state concentrations:
\begin{equation}
\Delta G^{o\ddag}_{new} = \Delta G^{o\ddag} + k_B  \ln \left (\frac{c^{0\ddag}}{c^{0\ddag}_{new}} \right) \label{eqDGnew}
\end{equation}
\eqname\ref{eqDGnew} confirms the intuition that more concentrated standard states have higher entropy and thus lower Gibbs free energy. Furthermore, substituting \eqname\ref{eqDGnew} into \eqname\ref{eqkconst} confirms that the value of the rate constant $k$, and thus also the rates in \eqnames\ref{eqr1d} and \ref{eqr1pr}, do not depend on the arbitrary choice of standard state concentration for the activated complex.

Going back to the rates in \eqnames\ref{eqr1d} and \ref{eqr1pr}, the activity products of the reactants in dissolution and precipitation reactions, $Q_{r,d}$ and $Q_{r,pr}$, can be defined more strictly than in the main body of this article. Since solids have unit activity, only the molecules in solution contribute to $Q\neq 1$, and these molecules are assigned to the 'background' in MASKE. More precisely, for the generic chemical reaction associated to the $allpar$ deletion or nucleation process, all the background species $i$ featuring in the reaction with negative stoichiometric coefficient (\ie consumed by the reaction) contribute to the activity product:
\begin{equation}
Q_r = \prod_{i\in bkg\;,\; \nu_i < 0} (\gamma_i c_i)^{-\nu_i} \label{eqQrApp}
\end{equation}
where $c_i$ is the concentration in solution of the molecular species, $\nu_i$ its stoichiometric coefficient in the reaction, and $\gamma_i$ its activity coefficient. If a molecule is labelled as $solvent$ in the $sol\_start$ command, the molecule is not included in the calculation of $Q_r$. 

MASKE computes $\gamma_i$ using the WATEQ Debye-H\"uckel formula discussed in the main body of this article, in \eqname\ref{eqDebH}.
If the user does not know the value of $a_i^o$ required there, by inputting a negative value in its place MASKE will use the Davies equation instead:
\begin{equation}
log_{10} (\gamma_i) = -z_i^2 A \left( \frac{\sqrt{I}}{1+ \sqrt{I}} -0.3 I \right)  \label{eqDavies}
\end{equation}
For uncharged species, with $z_i = 0$, \eqname\ref{eqDebH} reduces to Setchenow equation $log_{10} (\gamma_i) =  b_i I$ \citep{langmuir1997aqueous}
; if the user does not know the value of $b_i$, inputting a negative value in the $ChemDB$ file signals MASKE to employ $b_i = 0.1$ by default.

\begin{figure}[h]
\centerline{\includegraphics[width=0.99\textwidth] {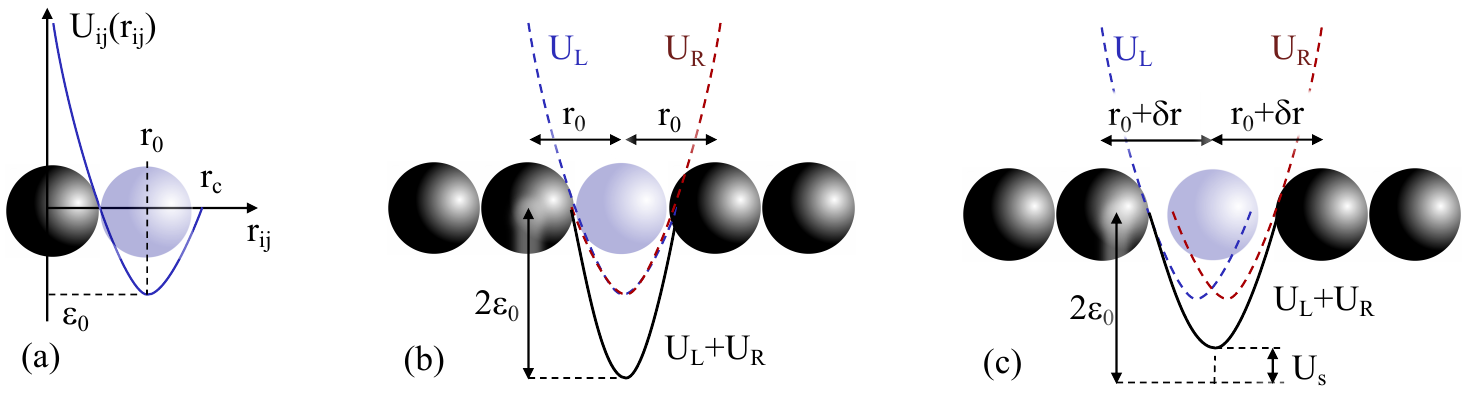}}
\caption{Changes in interaction energy accompanying the deletion or nucleation of a new particle, in blue, within a 1D array of other solid particles. (a) A typical pairwise potential of interaction between solid particles, here harmonic with cutoff. In the 1D array in (b) and (c), the change in interaction energy accompanying the deletion or nucleation of the blue particle comes from two contributions: $U_L$ for the interaction with the black particle on the left, and $U_R$ with the black particle on the right. (b) Energy change from particle deletion or nucleation when the sampled particle perfectly fits the 1D array, and (c) when it does not fit the array perfectly hence it experiences a strain energy $U_s$.}
\label{figDU}
\end{figure}

The contribution of mechanical interactions to the reaction rates in \eqnames\ref{eqr1d} and \ref{eqr1pr} come from their excess enthalpy terms, and specifically by the changes in total interaction energy $\Delta U_d$ and $\Delta U_{pr}$ caused by the dissolution or nucleation of the particle being sampled. \figname\ref{figDU} shows how, for pairwise interactions, these energy changes can be split into two contributions: a bond term $\varepsilon_0$ which sets the local minimum between two neighbouring particles at their equilibrium distance $r_0$, and a strain energy term $U_s$ that emerges when the local minimum resulting from multiple interactions leaves the sampled particle away from its equilibrium distance $r_0$ from its neighbours: see \figname\ref{figDU}c. For the example in \figname\ref{figDU}, therefore, if the blue particle is being sampled for deletion, its $\Delta U_d$ would be $2|\varepsilon_0|-U_s$ for the scenario in \figname\ref{figDU}c, whereas for nucleation $\Delta U_{pr} = -2|\varepsilon_0|+U_s$. As expected, a strong bond energy term (\viz a large, negative $\varepsilon_0$) leads to a large negative $\Delta U_{pr}$, which increases the precipitation rate in \eqname\ref{eqr1pr}, and to a large positive $\Delta U_d$, which decreases the dissolution rate in \eqname\ref{eqr1pr}. The opposite effect on rates comes from a large strain energy $U_s$, due to the local particle arrangement, which leads to a less negative $\Delta U_{pr}$ and to a less positive $\Delta U_d$.

\begin{figure}[h]
\centerline{\includegraphics[width=0.45\textwidth] {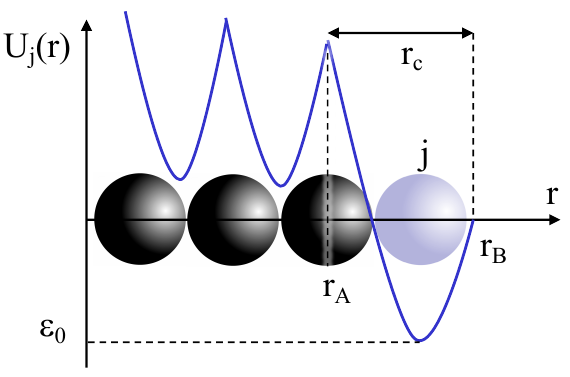}}
\caption{Energy change associated to the deletion or nucleation of particle $j$, in blue, depending on its position $r$ in the semi-infinite 1D array (\ie with no other particles on the right).}
\label{figBasin}
\end{figure}

\figname\ref{figDU} assumes that the sampled particle, in blue, sits in a local energy minimum resulting from the interactions with all its neighbours. In reality, if small nanoparticles are considered, thermal fluctuations will make the particle oscillate around its local energy minimum. For example, \figname\ref{figBasin} shows the energy landscape for particle $j$ depending on its location $r$ within a simple system consisting of a 1D array of other particles and going from $r=-\infty$ to $r=r_A$. Various local minima exist for particle $j$, the rightmost being the most energetically favourable in this case. This rightmost basin extends from $r_A$ to $r_B = r_A + r_c$, where $r_c$ is the interaction cutoff distance from \figname\ref{figDU}a. To account for thermal oscillations within a local energy basin, the rates in \eqnames\ref{eqr1d} and \ref{eqr1pr} should feature interaction energies that are averaged over such oscillations (\ie over all the possible configurations where the particle sits within the local energy basin, weighted by the corresponding Boltzmann factors). 

The $allpar$ mechanism in MASKE uses the simplifying assumption that the local energy minimum dominates the average interaction, which is reasonable for strong bond energies $\varepsilon_0$. This, however, means that the particles sampled for deletion or nucleation should occupy such local energy minima before computing the rates in \eqnames\ref{eqr1d} and \ref{eqr1pr}. When particle deletion is sampled, this requirement is not automatically satisfied by MASKE, but the user can easily ensure it by specifying a process of type $Every$ that performs an energy minimization in LAMMPS every time a continuous or discrete process is carried out. When instead nucleation is sampled, MASKE automatically searches for local energy minima by creating $M$ trial particles on a user-specified lattice and by running an energy minimization for them. This energy minimization requires each trial particle to individually find a local minimum: none of the current minimization styles in LAMMPS allows for this convergence condition to be imposed, hence an additional, bespoke LAMMPS minimization style called $quickmaske$ is included in the MASKE distribution (it uses the same algorithm as the $quickmin$ style in LAMMPS \cite{}). During this minimization, all the particles in MASKE are fixed to their positions except the $M$ trial particles being sampled for nucleation. Furthermore, the only active interactions are those between the $M$ sampled trial particles and the other particles that already exist in MASKE, with $real\_types$ in the input script; all the interaction between trial particles are instead turned off. As a result, the $M$ trial particles effectively sample the energy landscape of a new particle nucleating at any generic position within a user-specified region, which may well be the entire simulation box. The resolution of this sampling depends on how small the cell of the user-specified lattice is; $\Delta V$ in the precipitation rate in \eqname\ref{eqr1pr} is the volume of such lattice cell. For the example in \figname\ref{figBasin}, assume that a 1D lattice with $M$ positions defining small cells of size $\Delta r$ is used to sample possible nucleation sites between $r=0$ and $r_B$. A number $m$ of such positions will fall within the $A-B$ basin, \ie between $r_A$ and $r_B$. When running energy minimization on the trial particles, these $m$ particles will move to the local energy minimum, where the blue particle $j$ sits in \figname\ref{figBasin}. As a result, the nucleation of a particle in that energy minimum carries a weight equal to $m \Delta r$ or, for a more general 3D case, $m \Delta V = V_t$. This tributary volume $V_t$ is the volume of the energy basin, \viz the integral of all the locations $r$ pertaining to the basin (if sampled through infinitely small $\Delta V$). For the 1D case in \figname\ref{figBasin}, $V_t$ would be simply $r_B-r_A$.

The tributary volume of the local energy basin being sampled, $V_t$, appears in both the dissolution and the precipitation rates in \eqnames\ref{eqr1d} and \ref{eqr1pr}. In the dissolution rate, $V_t$ is raised to the power of $\alpha/3$, where $\alpha$ is the spatial dimensionality of $c^{o\ddag}_d$, which in turn matches the spatial dimensionality of the rate constants measured during the experiments that are used to obtain $\Delta G^{o\ddag}_d$. The $\alpha/3$ exponent ensures that the event-specific rate $r_{1,d}$ is overall in units of number of events per unit time. Dissolution and precipitation rate constants are typically expressed per unit surface, in which case $\alpha = 2$. The precipitation rate in \eqname\ref{eqr1pr} features a $\Delta V \cdot V_t^{\alpha/3-1}$ term instead. However, when all the $m$ trial particles pertaining to the same local energy basin are energy-minimised to the same local minimum, the cumulative rate of all these identical nucleation events (and thus the probability of selecting that site for nucleation, from \eqname\ref{eqPi}) ends up scaling as $m\Delta V \cdot V_t^{\alpha/3-1} = V_t \cdot V_t^{\alpha/3-1} = V_t^{\alpha/3}$, which brings back the same factor as in the dissolution rate. 

There is no easy way to compute $V_t$, which usually changes from one local basin to another. For example, the energetically favourable, rightmost basin in \figname\ref{figBasin} has $V_t = r_c$ (in 1D), the less energetically favourable basins to its left have $V_t = r_0 < r_c$, whereas the basins associated to the blue particle in \figname\ref{figDU} have $V_t = 2r_0$ and $V_t = 2(r_0 + \delta r)$ respectively in the stress-free system in \figname\ref{figDU}b and in the stressed system in in \figname\ref{figDU}c. Amorphous systems in 3D may also feature rugged energy landscapes, with narrow energy basins with $V_t < r_0$. The $allpar$ mechanism in MASKE works under the approximation that one same representative value of $V_t$ can be attributed to all basins; the user provides this value through scaling parameters attached to each $molecule$ defined in the $ChemDB$ file. As a result, for a particle whose deletion or nucleation features a given set of molecules in the foreground of its corresponding chemical reaction, $V_t$ is obtained as the sum of the scaled volumes of all these foreground molecules in the particle. For example, for the rightmost basin in \figname\ref{figBasin}, the user may provide a scaling factor of $r_c/r_0$ to the molecular volumes, since $r_0$ determines the volume that a particle in the array occupies, and $r_c$ is the size of the basin. In a typical dissolution/precipitation process, if the interaction potentials are single-well and relatively short-range, such as in \figname\ref{figDU}a, the most important events for the overall kinetics are particle deletions and nucleations on existing crystal surfaces, for which the scenario of the rightmost basin in \figname\ref{figBasin}, and thus the $r_c/r_0$ correction, are still relevant. This is because the volume that a particle occupies in a 3D crystal structure is still controlled by $r_0$, except that the interaction potential extends to $r_c$ in the direction perpendicular to the surface.

For the $\Delta V \cdot V_t^{\alpha/3-1}$ term in \eqname\ref{eqr1pr} to return the desired $V_t^{\alpha/3}$ weight factor after summing it over trial particles, the user must employ a sufficiently fine lattice, with small $\Delta V$, which may require a large number of trial particles $M$ to sample the nucleation region. For crystalline systems, the size of typical energy basins is in the range of a particle volume (\eg in \figname\ref{figBasin}) and the results in the main body of this article have shown that $\Delta V \approx V_p/8$ already provide a satisfactory sampling of nucleation sites.  

For the excess enthalpy in the $allpar$ mechanism, \eqname\ref{eqeps0} in the the main body of this articles has shown that a relationship must hold between the interaction energy of two bonded particles in stress-free conditions, $\varepsilon_0$, and the solid-solution interfacial energy of a particle $\gamma\Omega$. Namely:
\begin{equation}
\varepsilon_0 =  \frac{2 \gamma \Omega}{n_b}  \label{eqeps0App}
\end{equation}
where $n_b$ is the number of neighbours for a bulk particle. The user can manipulate the excess enthalpy term through their input of $\gamma$ in the $ChemDB$ file, and through their choice and parametrization of the interaction potentials in LAMMPS (\via the MASKE input script). This can lead to a wide range of different system behaviours. For example, by setting $\gamma = 0$ and an interaction potential where the bond energy $\varepsilon_0$ is set to zero as well, the user would create a simulation where particles would have the same probability of being deleted or nucleated irrespective of whether they sit next to other existing particles, or they are isolated in the solution, or even they completely overlap with other existing particles. This scenario is unphysical, hence the user must be careful in setting a meaningful relationship between $\gamma$ and interaction potentials when preparing a simulation. \eqname\ref{eqeps0App} is valid only for particles interacting through single-well potentials and that can form an orderly lattice. Modifications of the equation or of the definition of excess enthalpy altogether are needed, for example when dealing with amorphous agglomerates of solid particles that feature eigenstresses \cite{alex2023carbonation}, or with larger particles  that cannot be properly treated with the $allpar$ mechanism.

The last term discussed in relation to the the rate equations, \eqnames\ref{eqr1d} and \ref{eqr1pr}, is the fraction $\chi$, of excess enthalpy affecting the precipitation rate. If $\chi$ is set to zero, the excess enthalpy from solid-solid and solid-solution interactions only affects the dissolution rate. This is the usual assumption in Transition State Theory, where one assumes that the excess enthalpy raises the free energy level of the solid phases involved in the reaction and it is already fully lost in the activated complex state. As a result, with $\chi=0$ the excess enthalpy has no effect on the free energy barrier for precipitation, which involves solvated molecules as reactants and the activated complex. As noted in \cite{shvab2017precipitation}, however, $\chi = 0$ entails that particle nucleation becomes position independent (under the current solution model in MASKE, where ion concentrations and temperature are assumed uniform everywhere in a simulation box). This leads to the computationally inefficient scenario whereby many nucleation events may be carried out in energetically very unfavourable locations, \eg overlapping with other existing particles, just for the opposite deletion event to be carried out immediately afterwards due to the high excess enthalpy of the recently nucleated particle. 
Setting $\chi>0$ coincides with assuming that a fraction of the excess enthalpy is still present in the activated complex. If a dissolution event is being sampled and the excess enthalpy change is positive, \ie the event is enthalpy-wise unfavourable (\eg dissolving a well-connected particle in the bulk of a crystal), using $\chi>0$ reduces the activation energy for dissolution, thus increasing the dissolution rate. On the other hand, the opposite event of nucleating the same particle at the same site will also feature a lower activation energy compared to the $\chi=0$ case, because the activated complex now has a lower free energy due to its portion of favourable, \ie negative, free energy on the way towards the solid state. Hence also the nucleation rate back to the same position will increase. The increase of both dissolution and precipitation rates produces a compensation effect on the probability of a particle actually existing in that site, with more likelihood of seeing a particle dissolving and reforming there compared to the $\chi = 0$ case. The case of a deletion event with negative change in enthalpy instead, \ie an event that is enthalpy-wise favourable (\eg dissolving an particle under stress), will lead to a higher activation energy for dissolution and for precipitation back in the same site. As a result both dissolution and nucleation rates will decrease, once again producing a compensation effect on the probability of a particle existing in that site, but with less likelihood of seeing a particle dissolving and reforming there compared to the $\chi = 0$ case. This mitigates the aforementioned issue of nucleating particles in sites that are so energetically unfavourable that a subsequent deletion event would immediately remove them.

%%*****************************************
\section{Net rates equations in $allpar$, and recovering Transition State Theory}\label{secAppNetTST}
%%*****************************************

In the main body of this articles, the expressions of reaction rates for dissolution and precipitation, \eqnames\ref{eqr1d} and \ref{eqr1pr} (as per $allpar$ mechanism in MASTKE), have been used to also derive net reaction rates, in \eqnames\ref{eqrNd} and \ref{eqrNpr}. A few considerations underpin the derivation of the net rate expressions. Starting with the net dissolution rate in \eqname\ref{eqrNd}, a first consideration is that in the backward reaction, \viz straight precipitation, all the trial particles sampling precipitation into the same energy basin will have identical straight rate $r_{1,pr}$. This leads to the $V_t^{\alpha/3}$ term in \eqname\ref{eqrNd}, which comes from summing all the $\Delta V_t^{\alpha/3-1}$ terms from $r_{1,pr}$ pertaining to the same energy basin. Another consideration is the assumption that the activated state is identical for the forward (here, dissolution) and backward (here, precipitation) reactions, hence $c^{o\ddag}_d = c^{o\ddag}_{pr}$ and $\gamma^\ddag_d=\gamma^\ddag_{pr}$. Since precipitation in this case is simply the reverse reaction of dissolution, $n_{r,j,diss} = n_{r,j,nuc}$ and $Q_{r,pr}$, which is the activity product of the reactants in the precipitation reaction, has been re-labelled as $Q_{p,d}$, \ie the activity product of the products of the dissolution reaction. Lastly, \eqname\ref{eqrNd} features the equilibrium constant of the dissolution reaction, $K_{eq,d} = \left(\frac{Q_{p,d}}{Q_{r,d}}\right)_{eq}$, where the $eq$ subscript indicates that the activity products are measured at equilibrium. $K_{eq,d}$ appears in the rate equation because of its relationship to the standard change in free energy for the dissolution reaction, $\Delta G^o_d$, \ie $K_{eq,d} = \exp\left( -\frac{\Delta G^o_d}{k_BT}\right)$. This $\Delta G^o_d$ is the difference in standard free energy between products and reactants in the dissolution reaction, hence it determines the difference between forward and backward standard activation energies: $\Delta G^{o\ddag}_{pr} + \Delta G^{o}_{d} = \Delta G^{o\ddag}_{d}$. 

All the terms in the net dissolution rate expression, in \eqname\ref{eqrNd}, ultimately relate only to quantities that are defined by the user when inputting the parameters in the $ChemDB$ file, which that inform the chemical reactions associated to the dissolution itself. In evaluating the net dissolution rate, MASKE does not explicit sampling any nucleation event. Therefore, a simulation may sample exclusively particle deletion events, knowing that \eqname\ref{eqrNd} will reduce the dissolution rate by the rate of the corresponding but reverse precipitation reactions. A limitation of \eqname\ref{eqrNd} is that, if the contribution from the  precipitation rate is greater than from the dissolution rate, then $r_{1,d}^{net}$ will be zero and the dissolution event will never take place. This excludes the possibility of any energetically unfavourable fluctuation, which limits the ability to simulate some mechanisms such as crystal nucleation. This limitation has been addressed in the main body of this article.

Similar considerations as above apply to the derivation of the net precipitation rate, in \eqname\ref{eqrNpr}. There the backward reaction, which is a straight dissolution, refers to a corresponding precipitation event that covers only a portion $\Delta V$ of an energy basin with volume $V_t$. Hence $r_{1,d}$ carries the same $\Delta V \cdot V_t^{\alpha/3-1}$ as the precipitation event which is actually being sampled. Indeed, all term in \eqname\ref{eqrNpr} ultimately only refer to a precipitation reaction being sampled, similar to how all terms in \eqname\ref{eqrNd} only referred to the dissolution reaction. $Q_{p,pr}$ is the activity product of the products of the precipitation reaction, which replaces the activity product of the reactants of the dissolution reaction that, here, is just the reverse of the precipitation one. The equilibrium constant of the precipitation reaction, $K_{eq,pr} = \left(\frac{Q_{p,pr}}{Q_{r,p}}\right)_{eq}$ emerges in a similar way as $K_{eq,d}$ in \eqname\ref{eqrNd}. Namely, $\Delta G^o_{pr}$ is the standard change in free energy for the precipitation reaction, $K_{eq,pr} = \exp\left( -\frac{\Delta G^o_{pr}}{k_BT}\right)$ and $\Delta G^{o\ddag}_{d} + \Delta G^{o}_{pr} = \Delta G^{o\ddag}_{pr}$.

The net rate equations developed in this article are extensions of Transition State Theory (TST). TST ultimately leads to simple equations of net rate of the form \cite{lasaga2014kinetic}:
\begin{equation}
\mathcal{R}_{1,d}^{TST} = k (1-\beta) \label{eqrTSTd}
\end{equation}
$\mathcal{R}$ is the net rate of reaction per unit area or per unit volume, $k$ is the rate constant for the dissolution reaction, and $\beta = \frac{Q_{p,d}/Q_{r,d}}{K_{eq,d}}$ is the saturation index of the dissolution reaction. For solid dissolution, since a stress-free solid is commonly assumed to be in its standard state, $Q_{r,d} = 1$ and thus $\beta$ simplifies to $\frac{Q_{p,d}}{K_{eq,d}}$ or, that is the same if one considers the precipitation reaction opposite to the dissolution one, $\beta = \frac{Q_{r,pr}}{K_{eq,d}}$. The TST rate equation for the opposite precipitation reaction is obtained simply by detailed balance:
\begin{equation}
\mathcal{R}_{1,pr}^{TST} = k (\beta - 1) \label{eqrTSTpr}
\end{equation}
where $k$ and $\beta$ are still those for the dissolution reaction.
Under a set of circumstances, the TST rates in \eqnames\ref{eqrTSTd} and \ref{eqrTSTpr} are  recovered from the $allpar$ rates in MASKE, \viz from \eqnames\ref{eqrNd} and \ref{eqrNpr}. First of all, take $\chi = 0$, meaning that any excess enthalpy from solid-solid and solid-solution interactions only exists in the solid and is fully lost already in the activated complex state. Let us then define:
\begin{equation}
k = \kappa \frac{k_BT}{h}  \frac{c^{o\ddag}_d}{\gamma^\ddag_d} \exp\left( \frac{-\Delta G^{o\ddag}_{d}}{k_BT} \right) \label{eqk}
\end{equation}
which has units of number of events per unit time, per unit area or volume (depending on the dimensionality $\alpha$ of $c^\ddag$). With this one can already rewrite the non-zero case in \eqname\ref{eqrNd} as:
\begin{equation}
\mathcal{R}_{1,d}= \frac{r_{1,d}^{net}}{V_t^{\alpha/3}} = k
\left\{  \exp\left[ -\frac{\Delta U_{d}-\gamma \Omega}{k_BT} \cdot \frac{1}{n_{r,j,diss}} \right] Q_{r,d} - \frac{Q_{p,d}}{K_{eq,d}} \right\}  \label{eqrNdderiv}
\end{equation}
If the physically meaningful relationship between bond strength and surface energy in \eqname\ref{eqeps0} is imposed, the excess enthalpy $\Delta U_d - \gamma \Omega$  in \eqname\ref{eqrNdderiv} is zero for a stress-free particle occupying a kink position in its crystal lattice. Noting then that $Q_{r,d}=1$ for a dissolving solid and that $\frac{Q_{p,d}}{K_{eq,d}}$ is indeed the definition of $\beta$, one recovers indeed the TST net rate equation in \eqname\ref{eqrTSTd}. This means that the rate equation in MASKE is a more general form of \eqname\ref{eqrTSTd}, which reduces to the usual TST form for dissolution processes that are controlled by stress-free kink particles, if one assumes that excess enthalpy can only be present in the non-activated solid state. At the macroscopic level reaction rates are often expressed following \eqname\ref{eqrTSTd}; despite tempting, it would be incorrect to define a macroscopic rate constant $k$ as per \eqname\ref{eqk}. Indeed, a macroscopic rate constant implicitly accounts for morphological effect, such as for example the number of kink sites per unit area in a kink-controlled dissolution process, or for complex series of reactions all subsumed into a single, effective one \citep{cefis2017chemo}. \eqname\ref{eqk} instead applies only to a one-step reaction and does not subsume any morphological effect, as in the $allpar$ mechanism in MASKE they are accounted for through the excess enthalpy term in \eqname\ref{eqrNd}.

The TST precipitation rate in \eqname\ref{eqrTSTpr} is also a special case of MASKE's rate from \eqname\ref{eqrNpr}. To show this, let us first recall the common assumptions that $c^{o\ddag}_{pr}=c^{o\ddag}_{pr}$ and $\gamma^\ddag_{pr}=\ddag_d$, and the following relationships:  $\Delta G^{o\ddag}_{pr} + \Delta G^{o}_{d} = \Delta G^{o\ddag}_{d}$ and $K_{eq,d} = \exp\left( -\frac{\Delta G^o_d}{k_BT}\right)$. These lets us reformulate the prefactor in \eqname\ref{eqrNpr} (which one can identify as a precipitation rate constant $k_{pr}$) as:
\begin{equation}
k_{pr} = \kappa \frac{k_BT}{h}  \frac{c^{o\ddag}_{pr}}{\gamma^\ddag_{pr}} \exp\left( \frac{-\Delta G^{o\ddag}_{pr}}{k_BT} \right) = k   \exp\left( \frac{-\Delta G^{o}_{d}}{k_BT} \right) =  \frac{k}{K_{eq,d}}\label{eqkpr}
\end{equation}
where $k$ is the dissolution rate constant in \eqname\ref{eqk}. In fact, \eqname\ref{eqkpr} simply states the well-known equilibrium condition whereby $\frac{k}{k_{pr}} = K_{eq,d}$. If one assumes again $\chi = 0$ and stress-free kink particles with no excess enthalpy (from \eqname\ref{eqeps0}), \eqname\ref{eqrNpr} becomes:
\begin{equation}
\mathcal{R}_{1,pr}= \frac{r_{1,pr}^{net}}{\Delta V \cdot V_t^{\alpha/3-1}} = \frac{k}{K_{eq,d}}
\left( Q_{r,pr} - \frac{Q_{p,pr}}{K_{eq,pr}} \right)  \label{eqrNpderiv}
\end{equation}
Noting finally that $\frac{Q_{r,pr}}{K_{eq,d}}$ is the saturation index $\beta$ for the dissolution reaction, that $Q_{p,pr} = 1$ when the products are solids, and that $K_{eq,pr} = K_{eq,d}^{-1}$, one recovers the usual TST rate in \eqname\ref{eqrTSTpr}.

\end{document}